\documentclass[aps,a4paper, preprint, superscriptaddress,preprintnumbers,floatfix,nofootinbib,amsmath,amssymb]{revtex4-1}

\usepackage{url}
\usepackage{hyperref}
\usepackage{color}
\usepackage{cancel}
\usepackage{cleveref}
\usepackage{subfig}
\usepackage{soul}
\usepackage[normalem]{ulem}

\usepackage{amstext,amssymb}
\usepackage{amsmath}
\usepackage{graphicx}
\usepackage{diagbox}
\usepackage{slashbox}
\usepackage{slashbox}
\usepackage{xspace}
\usepackage{color}
\usepackage{units}
\usepackage[T1]{fontenc}
\usepackage{amsmath,bm}
\usepackage{nicefrac}
\usepackage{slashed} 
\usepackage{multirow}
\newcommand{\be}{\begin{equation}}
\newcommand{\ee}{\end{equation}}
\newcommand{\bea}{\begin{eqnarray}}
\newcommand{\eea}{\end{eqnarray}}

\usepackage{slashed}

\newcommand{\vem}{V_{e\mu}}
\newcommand{\vet}{V_{e\tau}}
\newcommand{\vmt}{V_{\mu\tau}}

\newcommand{\nua}[1]{\ensuremath{\rlap{\kern-2.5pt\ensuremath{\overset{\scriptscriptstyle(-)}{\phantom{\nu}}}}{\ensuremath{{\nu}_{#1}}}}\xspace}

\newcommand{\deltaCP}{\ensuremath{\delta_{\rm CP}}}

\definecolor{brickred}{rgb}{0.8, 0.25, 0.33}
\definecolor{brightcerulean}{rgb}{0.11, 0.67, 0.84}
\definecolor{brown(traditional)}{rgb}{0.59, 0.29, 0.0}
\begin{document}
\title{Study of Long Range Force in P2SO and T2HKK}
\author{Priya Mishra}
\email{mishpriya99@gmail.com}
\affiliation{School of Physics,  University of Hyderabad, Hyderabad - 500046,  India}
\author{Rudra Majhi}
\email{rudra.majhi95@gmail.com}
\affiliation{ Nabarangpur College, Nabarangpur - 764063, Odisha, India}
\author{Sambit Kumar Pusty}
\email{pustysambit@gmail.com}
\affiliation{School of Physics,  University of Hyderabad, Hyderabad - 500046,  India}
\author{Monojit Ghosh}
\email{mghosh@irb.hr}
\affiliation{Center of Excellence for Advanced Materials and Sensing Devices, Ru{\dj}er Bo\v{s}kovi\'c Institute, 10000 Zagreb, Croatia}
\author{Rukmani Mohanta}
\email{rmsp@uohyd.ac.in}
\affiliation{School of Physics,  University of Hyderabad, Hyderabad - 500046,  India}

\begin{abstract}

In this paper, we investigate the sensitivities of the upcoming long-baseline neutrino experiments P2SO and T2HKK to the long-range force (LRF). In the  context of these two experiments, our objective is to study (i) their capabilities  to put bounds on the LRF parameters as well as on constraining the mass and coupling strength of the new gauge boson, which  gives rise to LRF due to matter density in Sun and (ii) the effect of LRF in the measurement of standard oscillation parameters. In our study, we find that among the different neutrino experiments, the best bounds on the LRF parameters including mass of the new gauge boson and its coupling strength will come from the P2SO experiment. Our study also shows that LRF has non-trivial effect on the determination of the standard neutrino oscillation parameters except the precision of $\Delta m^2_{31}$. For this parameter, the precision remains unaltered in the presence of LRF for both these experiments.   
 
\end{abstract}

\maketitle
\flushbottom


\section{INTRODUCTION}

Neutrinos, being elusive subatomic particles with intriguing properties, hold great promise as a tool for discovering and understanding new phenomena in the realm of physics. Their unique characteristics, such as extremely weak interactions with matter and ability to change between different flavors, make them particularly interesting. Numerous experiments provide conclusive evidence that lepton flavor is not conserved during the propagation of neutrinos \cite{Esteban:2020cvm}. Neutrinos undergo oscillations, transitioning between different flavors, and the frequency of these oscillations is contingent upon the distance travelled and the energy of the neutrinos. The evolution of neutrino flavor during their traversal through matter is influenced by the Mikheev-Smirnov-Wolfenstein (MSW) mechanism \cite{Wolfenstein:1977ue}. This mechanism is characterized by the elastic forward scattering of neutrinos with matter.  In the framework of the Standard Model (SM) of particle interactions, this effect is precisely determined. The presence of electrons in matter creates an effective potential for the neutrinos, known as the matter potential ($V$),  plays a crucial role in influencing flavor transitions \cite{Barger:1980tf}. In the MSW mechanism, this potential is proportional to the electron number density at the neutrino's position:
\begin{equation}
    V_{CC}=\sqrt{2}G_F{N_e(r)}.
    \label{VCC}
\end{equation}
Here, $G_F$ is the Fermi constant, and $ N_e(r)$ is the number density of electrons. The matter potential can be modified by introducing flavor dependent interactions and such alterations have the capacity to reshape the pattern of flavor transitions experienced by neutrinos as they traverse through matter. One example of such interactions is flavor-dependent vector-like leptonic Long Range Force (LRF) \cite{Grifols:1993rs, Grifols:1996fk, Grifols:2003gy}.

The effect of LRF can be realized in general, through the $U(1)$ extension of the Standard Model, where the SM  gauge group $ SU(2)_L \times U(1)_Y$,  augmented by an additional $ U(1)_X$ symmetry. The anomalies associated with  $U(1)_{L_e}$, $U(1)_{L_\mu}$,  $U(1)_{L_\tau}$ and $U(1)_{B/3}$ symmetries  are equivalent, and hence combinations such as $U(1)_{L_e-L_\mu}$, $U(1)_{L_e-L_\tau}$ and $U(1)_{L_\mu-L_\tau}$ can be consistently accommodated in a way that maintains the gauge group as anomaly free \cite{Foot:1990mn,He:1990pn, He:1991qd, Foot:1994vd}.  It is important to emphasize that at any given moment, any one of these three symmetries can be seamlessly incorporated into the SM gauge group, as outlined in Refs. \cite{He:1990pn, He:1991qd}. This paves the way for them to be potentially natural and efficient extensions of the SM. An additional  new neutral vector boson is introduced by gauging each one of these symmetries, which mediates novel neutrino interactions with matter. In the context of local symmetry, the Goldstone boson is absorbed by the corresponding gauge boson, resulting in the acquisition of mass by the latter. The mass scale of this gauge boson can be interpreted in two ways: either through contact interaction or long range interaction.  If the new gauge boson possesses sufficient mass, it will induce contact interactions and consequently lead to the emergence of Neutral Current (NC) Non-Standard Interactions (NSI) \cite{Farzan:2017xzy}. On the flip side, if the mediator is excessively lightweight, the approximation of contact interaction becomes invalid, and the flavor-specific forces between neutrinos and matter particles extend over long distances. In such instances, neutrino propagation can still be characterized in terms of a matter potential. However, this potential is no longer solely determined by the particle number density at the neutrino's position in the medium. Instead, it relies on the average matter density within a radius of approximately $1/M_{Z^\prime}$ around the neutrino with $M_{Z^\prime}$ being the mass of the new gauge boson $Z^\prime$ \cite{Smirnov:2019cae}. With such an exceedingly small mass for $M_{Z^\prime}$, its interaction with matter is extremely weak, resulting in a very small gauge coupling ($g^\prime$). For this reason, this scenario can be considered as an invisible sector. There exist constraints on the LRF parameters expressed in terms of $\alpha_{\small {LRF}} = g^{\prime \hspace{0.5mm}2}/4\pi$ derived from the measurement of differential acceleration of the earth and moon towards the Sun from lunar ranging~\cite{Williams:1995nq,Dolgov:1999gk,Adelberger:2003zx,  Williams:2004qba}. The value of the LRF parameter has been obtained as $\alpha_{\small {LRF}}= 3.4 \times 10^{-49}$ at $2\sigma$ C.L. for the interaction range $\leq$ 1 A.U.

In general, interactions induced by LRF alter the matter potential in neutrino propagation. Because of this alteration of the matter potential, the probabilities of neutrino oscillation will get modified and therefore one can study the effect of LRF in  neutrino oscillation experiments. The concept of long range interaction with the help of flavor symmetries  in the context of neutrino oscillation was first introduced in Ref.~\cite{Joshipura:2003jh}, where the bounds on flavor dependent LRF parameters are obtained from atmospheric neutrinos. Later, bounds from solar neutrinos are also obtained in Ref.~\cite{Bandyopadhyay:2006uh}. In a study conducted in Ref.~\cite{Gonzalez-Garcia:2006vic}, bounds on the couplings for both vector and non-vector long-range forces have been reported from the oscillation of the solar neutrinos. There exist several other descriptions in the literature, for example in Refs.~\cite{Grifols:2003gy,Samanta:2010zh, Chatterjee:2015gta,Khatun:2018lzs,Singh:2023nek, Agarwalla:2023sng, Davoudiasl:2011sz}, the effect of  LRF in neutrino oscillation experiments has been illustrated. In this paper, we will study the effect of LRF in the upcoming long-baseline experiments P2SO \cite{Akindinov:2019flp} and T2HKK \cite{Hyper-Kamiokande:2016srs}. For the study of LRF in other two future long-baseline experiments i.e., T2HK \cite{Hyper-Kamiokande:2018ofw} and DUNE \cite{DUNE:2020ypp}, we refer to Refs.~\cite{Singh:2023nek,Chatterjee:2015gta}. In this study, our aim is to estimate the upper bounds of the LRF parameters and consequently study their effects  on the determination of the standard oscillation parameters. We will also estimate the bounds on the mass of the new gauge boson and the new gauge coupling for LRF due to the Sun.

The  paper is structured in the following manner. In the next section, we will describe the theoretical overview of the long-range force and write down the expressions for the LRF potentials.  Next, we will show how these potentials can be incorporated in the neutrino oscillation Hamiltonian in a model independent way. After this, we will write down a prescription on how one can calculate the neutrino oscillation probabilities in the presence of a LRF. In section \ref{exp}, we will provide the configurations of the experiments that we will use in our calculation. Section \ref{sim} outlines simulation details of our numerical analysis. In section \ref{res}, we present our results and finally, in section \ref{sum}, we summarize our findings and conclude. 


\section{Theoretical Background}
\label{Sec: Theory-BG}

According to the SM, lepton family number is considered to be conserved in nature. However, several decay processes such as $\mu \to e \gamma$, $\mu \to e e \overline{e}$, and $\tau \to \mu \gamma$ with upper limits on their respective branching ratios (BRs) as, $4.2 \times 10^{-13}$ \cite{MEG:2016leq}, $1 \times 10^{-12}$ \cite{SINDRUM:1987nra} and $4.2 \times 10^{-8}$ \cite{Belle:2021ysv}, are considered as potential indications of lepton flavor violation (LFV). As these BRs are relatively small, it is challenging to conclusively confirm the existence of LFV in nature. We take advantage of the small BRs of the LFV processes and assume that lepton flavor conservation could be a fundamental symmetry of nature and hence, the specific linear combinations involving $L_e$, $L_\mu$, and $L_\tau$ may be subject to gauging \cite{He:1990pn,He:1991qd,Foot:1990mn,Foot:1994vd}. In this work, we consider three such combinations which are: ${L_e-L_\mu}$, ${L_e-L_\tau}$ and ${L_\mu-L_\tau}$ symmetries. In scenarios, where the additional $U(1)$ corresponds to ${L_e-L_j}$ for $j=(\mu,\tau$), the electrons within the celestial bodies such as the Sun or the Earth create a potential that influences neutrinos in terrestrial experiments \cite{Joshipura:2003jh,Grifols:2003gy,Bandyopadhyay:2006uh}. The flavor-specific nature of $U(1)_{L_e-L_j}$ leads to alterations in neutrino oscillations, thus provides a pathway for constraining the gauge coupling. The effective potential for neutrinos on Earth, in such case is defined as,
\begin{equation}
     V_{ej} = g_{ej}^2 \frac{N_e}{4\pi r} e^{-M_{Z_{ej}}r},
\label{v_ej}
\end{equation}
where, $g_{ej}$ is the new gauge coupling i.e., $g_{e\mu}$ for ${L_e - L_\mu}$ and $g_{e\tau}$ for ${L_e - L_\tau}$ symmetries, $N_e \simeq 10^{57} $ \cite{bahcall1989neutrino}, is the number of electrons inside the Sun\footnote{The $e$, $p$, $n$ within the Earth can also generate a long-range potential at the neutrino site. However, it's noteworthy that this potential is approximately 10 times smaller compared to the potential generated by matter in the Sun \cite{Bandyopadhyay:2006uh}.}, $M_{Z_{ej}}$ is the mass of new gauge boson and $r$ measures the distance between the source of the potential and the neutrinos on Earth. 
Due to the absence of muons and taus on earth matter, it is not straight forward to investigate the effect  of ${L_\mu- L_\tau}$ symmetry, as its associated gauge boson $Z_{\mu\tau}$ does not directly couple to electrons, protons or neutrons. Nevertheless, there exists an indirect effect stemming from the kinetic mixing between the SM gauge boson $Z$ and $Z_{\mu\tau}$. Assuming that there are equal number of electrons and protons inside the Sun, so that their effect cancel each other completely and neutrinos on Earth are 
influenced by only neutrons inside the Sun. Such effective potential can be defined as \cite{Heeck:2010pg},
\begin{equation}
    \vmt = g_{\mu\tau} (\xi - \sin{\theta_w} \chi) \frac{e}{4\sin{\theta_w}\cos{\theta_w}}\frac{N_n}{4\pi r} e^{-M_{Z_{\mu\tau}}r},
    \label{v_mt}
\end{equation}
where, $g_{\mu\tau}$ is the new gauge coupling associated with the new gauge boson $Z_{\mu\tau}$ with mass $M_{Z_{\mu\tau}}$, $\chi$ is the kinetic mixing parameter between $Z$ and $Z_{\mu\tau}$ \cite{Holdom:1985ag}, $\xi$ is the rotation angle between mass and flavour bases of gauge bosons, $\theta_w$ is the Weinberg angle and  $N_n \simeq N_e/4$ $\simeq 1.5 \times 10^{56}$ \cite{Heeck:2010pg}, is the number of neutrons in the Sun. The bound on the value of  $(\xi - \sin{\theta_w} \chi)$ is $\leq 5 \times 10^{-24}$ \cite{Heeck:2010pg} and ${e}/{\sin{\theta_w}\cos{\theta_w}}$ = 0.723.

Our goal in this work is to study the effects of the LRF potentials $\vem$, $\vet$ and $\vmt$ in the neutrino oscillation experiments.


\section{Formalism}
\label{Sec: Form}

In the three flavor scenario, the effective Hamiltonian (in mass basis) for the propagation of neutrinos in the vacuum is given as follows:
\begin{eqnarray}
    H_{vac}= \frac{1}{2E} \begin{pmatrix}
         m^2_{1} & 0 & 0 \\
        0 &  m^2_{2} & 0 \\
        0&0&m^2_{3}
    \end{pmatrix},
\end{eqnarray}
where $m_1$, $m_2$, $m_3$ are the masses of the neutrinos ($\nu_1, \nu_2, \nu_3$) and $E$ is their energy. Including the matter potential and the potential due to LRF in flavor basis one obtains,
\begin{eqnarray}
    H_{\nu / \overline{\nu}}= \frac{1}{2E} \left [ U \begin{pmatrix}
         0 & 0 & 0 \\
        0 & \Delta m^2_{21} & 0 \\
        0&0&\Delta m^2_{31}
    \end{pmatrix}U^{\dagger} \right ] \pm H_{\rm matter} \pm H_{\small LRF},
    \label{Eqn: H_Total}
\end{eqnarray}
with
\begin{align}\label{lrf}
    H_{LRF} =\begin{cases}{\rm diag}(\vem, -\vem, ~0) ~~~~~~{\rm ~for}~ ~U(1)_{L_e - L_\mu},  \\ 
             {\rm diag}(\vet, ~ 0, -\vet) ~~~~~~{\rm ~for}~~   U(1)_{L_e - L_\tau}, \\ 
             {\rm diag}(0, \vmt, -\vmt) ~~~~~~~{\rm ~for}~ ~ U(1)_{L_\mu - L_\tau},
            \end{cases}
\end{align}
where ‘$+$’ sign is for neutrino and ‘$-$’ sign is for antineutrino flavor states, the unitary PMNS matrix $U$ contains three mixing angles, $\theta_{12}$, $\theta_{13}$ and $\theta_{23}$ as well as one Dirac-type CP phase $\delta_{\rm CP}$, which facilitate the rotation from the mass basis to the flavor basis,  and $\Delta m^2_{ij} = m_i^2 -m_j^2$ $(i=2, 3~$\&$~j = 1)$. Since matter contains only  $e$,  $u$ and $d$, the form of $H_{\rm matter}$ will have only ‘$ee$’ term and this will be equal to the earth matter potential  $V_{\rm CC}$ defined in Eqn.~\eqref{VCC}. The matter potential can be further simplified as,
\begin{align}
    V_{CC} \simeq  7.5 \hspace*{0.1 true cm} X_e \left (\frac{\rho}{10^{14} \hspace*{0.1 true cm}{\rm (g/cm^{3})}}\right )~ {\rm eV}\;,
\end{align}
where $X_e$ represents the ratio of the number of electrons $(N_e)$ to the sum of the number of protons $(N_p)$ and neutrons $(N_n)$. For a neutral medium, where the charges are balanced, $X_e$ is equal to 0.5. The value of $\rho$  is the earth matter density which is around $\rm 2.95 ~g/cm^3$ for the P2SO experiment and around $\rm 2.7 ~g/cm^3$ for the T2HKK experiment.

The structure of $H_{LRF}$ depends on the coupling of ${Z_{\alpha\beta}}$ to the lepton generations. For instance, as  $Z_{e\mu}$ can only couple to the first and second generations of leptons, the resulting potential emerges solely in the `$ee$' and `$\mu\mu$' terms, while other matrix elements are zero in $H_{LRF}$. 


\section{Analytical Expressions}

The analytical expressions for the oscillation probabilities in the presence of LRF for $L_{e}-L_{\mu}$ and $L_{e}-L_{\tau}$ cases have been derived in Ref.~\cite{Chatterjee:2015gta,Khatun:2018lzs} and for the   $L_{\mu}-L_{\tau}$ case, one can follow  the formalism described in Ref.~\cite{Agarwalla:2021zfr}. The total Hamiltonian $H_{\nu}$ defined in Eqn.~\eqref{Eqn: H_Total} can be diagonalized using a unitary matrix $\hat{U}$ defined as,
\begin{align}
    \hat{U} =  \hat{R}(\theta^\prime_{23})\hat{R}(\theta^\prime_{13})\hat{R}(\theta^\prime_{12})\;,
\end{align}
where for simplicity, it is assumed that CP is conserved.
 $\theta^\prime_{ij}$ are the modified mixing angles and can be expressed as,
\begin{align}
     \tan 2\theta^\prime_{23}&\simeq\frac{\Gamma\sin 2\theta_{23}-\zeta\cos2\theta_{23}}{\Gamma \cos2\theta_{23}+\zeta\sin2\theta_{23}-2\hat{B}} \;, \nonumber \\ 
     \tan2\theta^\prime_{13}&\simeq\frac{\sin2\theta_{13}(1-\alpha {\sin^2\theta_{12}})\cos\Delta\theta_{23}-\eta \sin\Delta \theta_{23}}{(\lambda_3-\hat{A}-\alpha \sin^2\theta_{12}\cos^2\theta_{13}-\sin^2\theta_{13})} \;,\nonumber \\
     {\tan2\theta^\prime_{12}}&\simeq
\frac{{\cos\theta^\prime_{13}}[\eta\cos\Delta\theta_{23}+\sin2\theta_{13}(1-\alpha \sin^2\theta_{12})\sin\Delta\theta_{23}]}{(\lambda_2-\lambda_1)} \;,
\end{align}
with
\begin{align}
&\lambda_1=\frac{1}{2}\Big[\lambda_3+\hat{A}+\sin^2\theta_{13}+\alpha \sin^2\theta_{12}\cos^2\theta_{13}
-\frac{\lambda_{3}-\hat{A}-\sin^2\theta_{13}-\alpha \sin^2\theta_{12}\cos^2\theta_{13}}{\cos2\theta^\prime_{13}}\Big] \;,\nonumber\\
&
\lambda_2=\frac{1}{2}\Big[\Gamma + 2\alpha \cos^2\theta_{12} -\frac{\zeta\sin2\theta_{23}+\Gamma\cos2\theta_{23}-2\hat{B}}{\cos2\theta^\prime_{23}}\Big] \;,
\nonumber\\
&\lambda_3= \frac{1}{2}\Big[\Gamma +2\alpha \cos^2\theta_{12} 
+\frac{\zeta\sin2\theta_{23}+\Gamma \cos2\theta_{23}-2\hat{B}}{\cos2\theta^\prime_{23}}\Big] \;,
\end{align}
and 
\begin{align}
 \Gamma = \cos^2\theta_{ 13}-\alpha \cos^2\theta_{12}+\alpha \sin^2\theta_{12}\sin^2\theta_{13} \;, \nonumber \\
 \zeta = \alpha\sin2\theta_{12}\sin\theta_{13}\;, ~~\eta = \alpha\sin2\theta_{12}\cos\theta_{13}\;,
 \end{align}
\begin{align}
    \alpha = \frac{\Delta m^2_{21}}{\Delta m^2_{31}}, \,~~ \hat{A}=\frac{2EV_{CC}}{\Delta m^2_{31}}, \,\,\, \hat{B} = \frac{2EV_{\mu\tau}}{\Delta m^2_{31}}\;,\,\,\, \Delta \theta_{23} = \theta_{23} - \theta^\prime_{23}\;.
\end{align}
 The diagonalised form of $H_{\nu}$ can be written as,
\begin{align}
    \hat{H}_{\nu}=\frac{1}{2E} \begin{pmatrix}
        m^{\prime 2}_{1} & 0 & 0\\
        0 & m^{\prime 2}_{2} & 0\\
        0& 0 & m^{\prime 2}_{3}
    \end{pmatrix}\;,
\end{align}
with $m^\prime_{i}$ ($i = 1,2,3$) being the modified masses of neutrinos defined as,
\begin{align}
m^{\prime 2}_{1} \simeq &\frac{\Delta m^2_{31}}{2}\Big[\lambda_1+\lambda_2+\frac{\lambda_1-\lambda_2}{\cos2\theta^\prime_{12}}\Big] \,,
\label{eq:m1}
\end{align}
\begin{align}
m^{\prime 2}_{2} \simeq &\frac{\Delta_{31}}{2}\Big[\lambda_1+\lambda_2-\frac{\lambda_1-\lambda_2}{\cos2\theta^\prime_{12}}\Big] \label{eq:m2} \,, 
\end{align}
\begin{align}
m^{\prime 2}_{3} \simeq &\frac{\Delta m^2_{31}}{2}\Big[\lambda_{3}+\hat{A}+\sin^2\theta_{13}+\alpha \sin^2\theta_{12}\cos^2\theta_{13}+\frac{\lambda_{3}-\hat{A}-\sin^2\theta_{13}-\alpha \sin^2\theta_{12}\cos^2\theta_{13}}{\cos2\theta^\prime_{13}}\Big] .\label{eq:m3}
\end{align}
The probability of $\nu_\mu \rightarrow \nu_e$ transition can be given as \cite{Chatterjee:2015gta,Agarwalla:2013tza},
\begin{align}
P_{\nu_\mu \rightarrow \nu_e}
 &=  \, \, 4\,\hat{U}_{\mu 2}^2 \hat{U}_{e 2}^2 \sin^2\frac{\Delta m^{\prime 2}_{21}L}{4E}
     +4\,\hat{U}_{\mu 3}^2 \hat{U}_{e 3}^2 \sin^2\frac{\Delta m^{\prime 2}_{31}L}{4E} \cr
   & + \, \,2\;\hat{U}_{\mu 3}\hat{U}_{e 3}\hat{U}_{\mu 2}\hat{U}_{e 2}
      \left(4\sin^2\frac{\Delta m^{\prime 2}_{21}L}{4E}\sin^2\frac{\Delta m^{\prime 2}_{31}L}{4E}\right) \cr
   & +\,\,2\;\hat{U}_{\mu 3}\hat{U}_{e 3}\hat{U}_{\mu 2}\hat{U}_{e 2}
      \left(\sin\frac{\Delta m^{\prime 2}_{21}L}{2E}\sin\frac{\Delta m^{\prime 2}_{31}L}{2E} \right)  \;,
\label{eq:prob-numu-to-nue}
\end{align}
and the expression for $\nu_\mu \rightarrow \nu_\mu$ survival probability can be given as, 
\begin{align}
P_{\nu_{\mu}\rightarrow\nu_{\mu}}
=  1 & - \left [4\,\hat{U}_{\mu 2}^2 \left(1 - \hat{U}_{\mu 2}^2 \right)
            \sin^2\frac{\Delta m^{\prime 2}_{21}L}{4E} \right. \nonumber \\
  & \left. \,\, + 4\,\hat{U}_{\mu 3}^2 \left(1 - \hat{U}_{\mu 3}^2 \right)
            \sin^2\frac{\Delta m^{\prime 2}_{31}L}{4E}  \right. \nonumber \\ 
 &   \left.
        -\,\, 2\, \hat{U}_{\mu 2}^2 \hat{U}_{\mu 3}^2
          \left(4\sin^2\frac{\Delta m^{\prime 2}_{21}L}{4E}\sin^2\frac{\Delta m^{\prime 2}_{31}L}{4E} \right) \right. \nonumber \\
 &  \left.
        - \,\, 2\, \hat{U}_{\mu 2}^2 \hat{U}_{\mu 3}^2
           \left(\sin\frac{\Delta m^{\prime 2}_{21}L}{2E}\sin\frac{\Delta m^{\prime 2}_{31}L}{2E} \right) \right]  \;,
\label{eq:prob-numu-to-numu}
\end{align}
 where, $L$ represents the baseline which is 2595 km for P2SO and 1100 km for T2HKK and $\Delta m^{\prime 2}_{ij}$ are the effective mass square differences.


\section{Experimental Details}
\label{exp}

In this study, we mainly focus on the two upcoming long-baseline experiments P2SO and T2HKK. Here are the relevant experimental details: 

\subsection{P2SO}

The Protvino to Super-ORCA (P2SO) is an impending long-baseline experiment where neutrinos will be produced at a U-70 synchrotron located at Protvino, Russia, and will propagate towards the detector situated at a distance of 2595 km in the Mediterranean Sea 40 km offshore Toulon, France. We refer to Refs.~\cite{Akindinov:2019flp, Singha:2022btw,Majhi:2022fed,Singha:2023set} for the detailed configuration of the P2SO experiment. The accelerator will produce a 450 KW beam corresponding to $4 \times 10^{20}$ protons on target annually for P2SO configuration. The Super-ORCA detector will use ten times more denser detector compared to the ORCA. Energy window ranges from 0.2 GeV to 10 GeV for P2SO experiment. We have considered total run period of six years consisting of three years in neutrino and three years in antineutrino modes. 


\subsection{T2HKK}
 In our  computational analysis for T2HKK experiment, we adhere to the configuration outlined in Ref \cite{Hyper-Kamiokande:2016srs}. The neutrino source is situated at J-PARC, featuring a beam power of 1.3 MW and a total exposure of $27 \times 10^{21}$ protons on target (POT), corresponding to a comprehensive 10-year operational period. This operational timeline is divided, with 2.5 years dedicated to neutrino mode and the subsequent 7.5 years to anti-neutrino mode, maintaining a balanced $1:3$ ratio between neutrinos  and anti-neutrinos. This setup will have a detector of 187 kt fiducial volume positioned at 295 km from the source, exposed to a  $2.5^\circ$ off-axis flux. Additionally, there will be another detector with the same volume, situated 1100 km from the source, encountering a  $1.5^\circ$ off-axis flux. The energy window under consideration spans from 0-3 GeV. This is an alternate setup of the T2HK (Tokai to Hyper Kamiokande) experiment where both the detectors will be placed at 295 km.

\section{Simulation Details}
\label{sim}

We have simulated P2SO and T2HKK experiments using GLoBES \cite{Huber:2004ka, Huber:2007ji} software package. In order to implement LRF we have modified the probability engine of GLoBES. We have estimated the sensitivity in terms of $\chi^2$ analysis, where we use the Poisson log-likelihood and assume that it is $\chi^2$-distributed:
\begin{equation}
 \chi^2_{{\rm stat}} = 2 \sum_{i=1}^n \bigg[ N^{{\rm test}}_i - N^{{\rm true}}_i - N^{{\rm true}}_i \log\bigg(\frac{N^{{\rm test}}_i}{N^{{\rm true}}_i}\bigg) \bigg]\,,
\end{equation}
where $N^{{\rm test}}$ and $N^{{\rm true}}$ are the number of events in the test and true spectra respectively, and $n$ is the number of energy bins. The systematic error is incorporated by the method of pull~\citep{Fogli:2002pt,Huber:2002mx}. The values of the oscillation parameters are taken from NuFit 5.2 and are listed in Tab.~\ref{table_sparam}.
\begin{table}[h] 
\centering
\begin{tabular}{|c|c|} \hline
Parameters            & True values $\pm$ $1\sigma$       \\ \hline
$\sin^2 \theta_{12}$  & $0.303^{+ 0.012}_{- 0.012}$      \\ 
$\sin^2 \theta_{13}$ & $0.02225^{+ 0.00056}_{- 0.00059}$                 \\ 
$\sin^2 \theta_{23} $ & $0.451^{+ 0.019}_{- 0.016}$                 \\ 
$\delta_{\rm CP}[^\circ] $  & $ 232^{+ 36}_{- 26}$         \\ 
$\Delta m^2_{21}$ [10$^{-5}$ eV$^2$]    & $7.41^{+ 0.21}_{- 0.20}$  \\ 
$\Delta m^2_{31}$ [10$^{-3}$ eV$^2$]   & $2.507^{+ 0.026}_{- 0.027}$    \\ 
 \hline
\end{tabular}
\caption{Values of oscillation parameters  with their 1$\sigma$ errors used for the study~\cite{Esteban:2020cvm}.}
\label{table_sparam}
\end{table}   
While calculating the $\chi^2$, the true values of the oscillation parameters are always kept at their best-fit values as shown in Tab.~\ref{table_sparam}. The relevant oscillation parameters are minimized in the test using the current uncertainties associated with these parameters. We will present all our results for the normal ordering of the neutrino masses i.e., $\Delta m^2_{31} > 0$. 


\section{Results}
\label{res}
In this section, we will present our results. First, we will show how the appearance and disappearance  probabilities depend on the LRF parameters. Then we will compute the bounds on the LRF parameters with P2SO and T2HKK. After that, we will study the effect of the LRF parameters in the determination of CP violation, octant of $\theta_{23}$ and mass ordering of the neutrinos for these two experiments. Then we will investigate whether the precision of $\deltaCP$, $\Delta m^2_{31}$ and $\theta_{23}$ will be affected if LRF exists in nature. Finally, we will show the capability of the aforementioned experiments to constrain the mass and coupling of the  new gauge boson for the long-range force due to Sun.

\begin{figure}
\begin{center}
    \includegraphics[width=80mm, height=70mm]{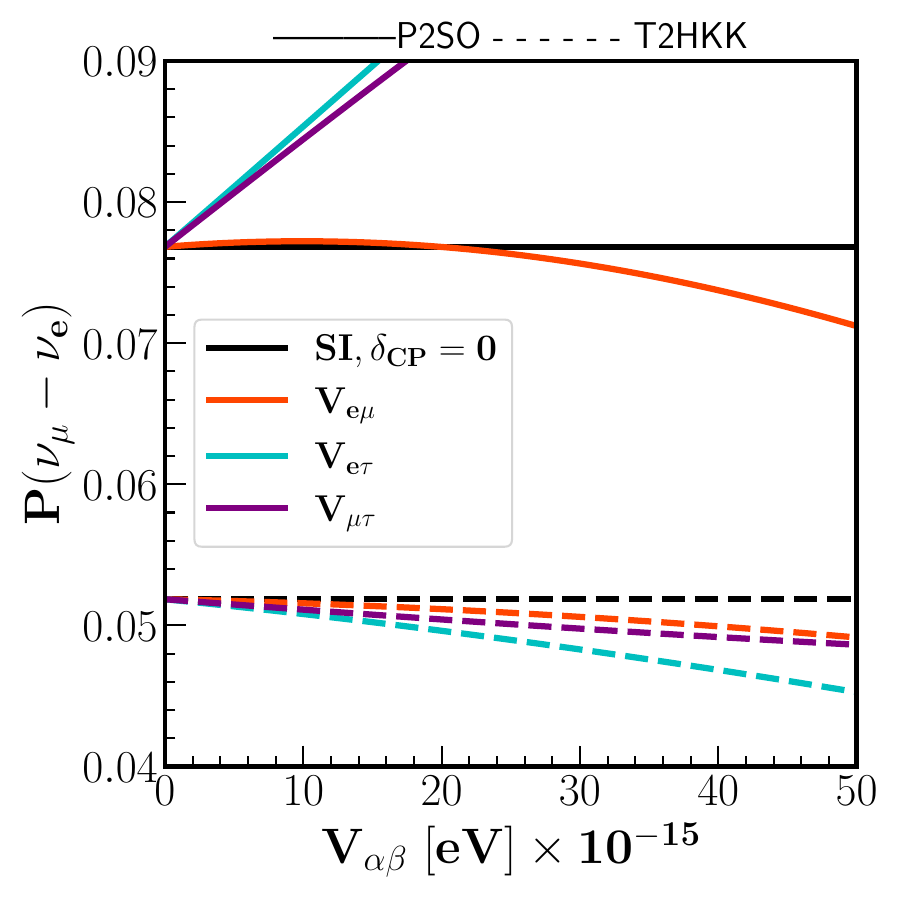}
    \includegraphics[width=80mm, height=70mm]{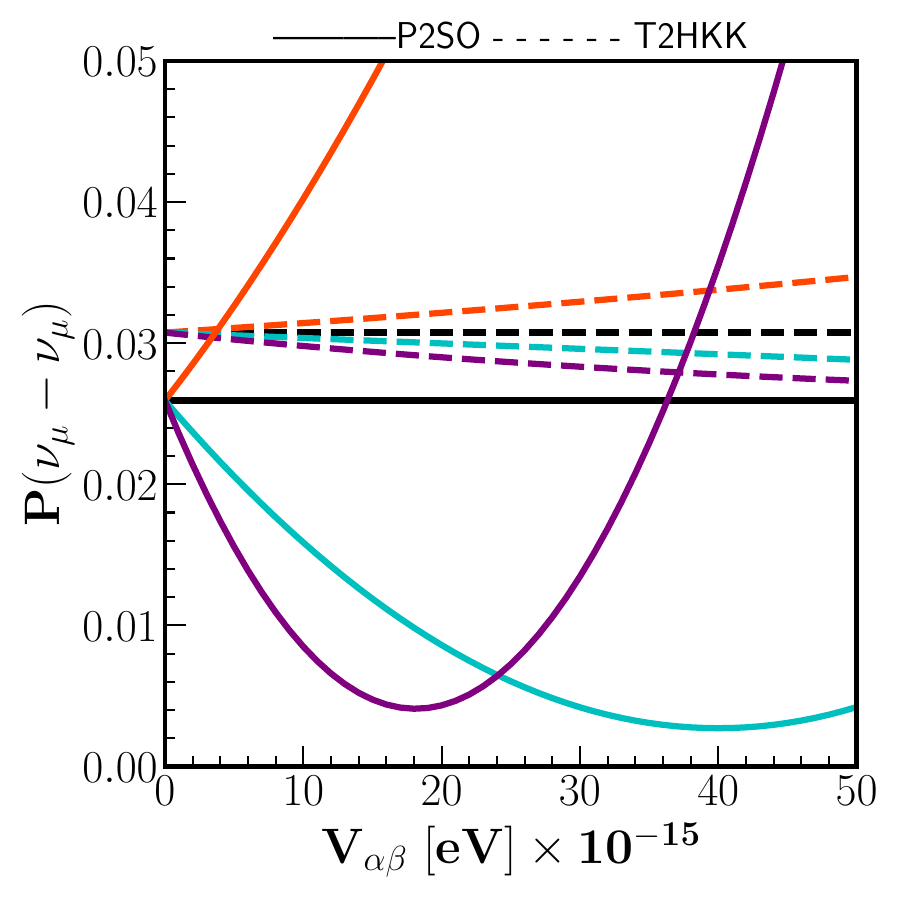}  \\
    \includegraphics[width=80mm, height=70mm]{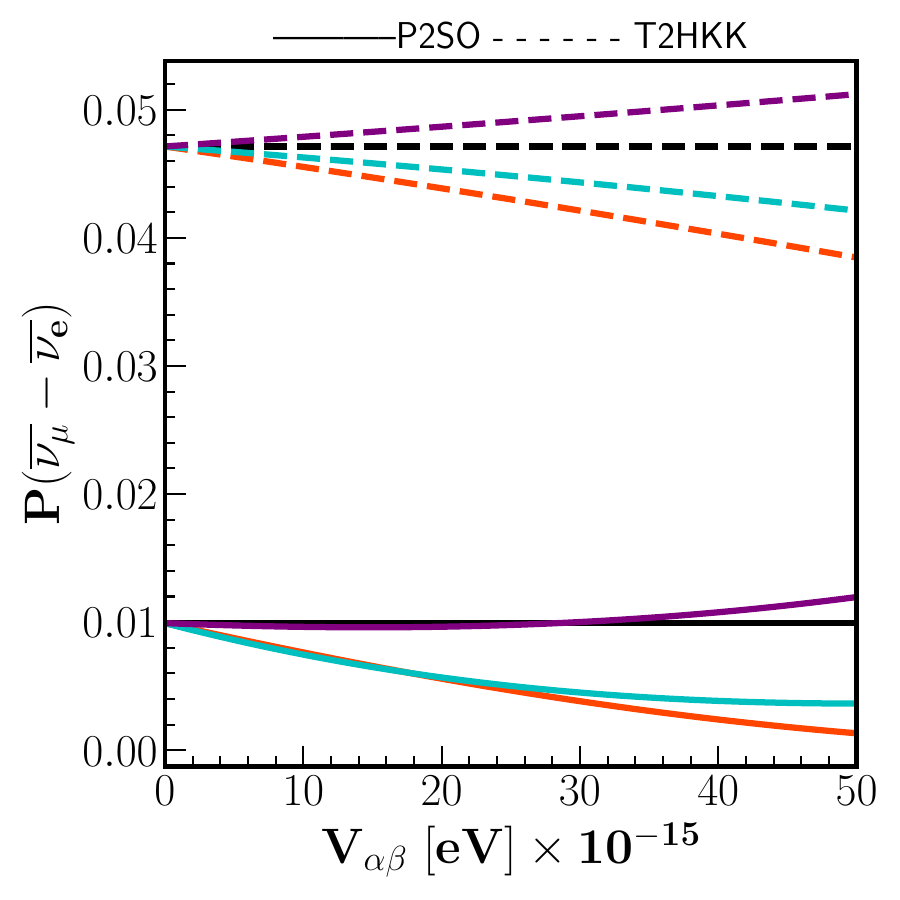}
    \includegraphics[width=80mm, height=70mm]{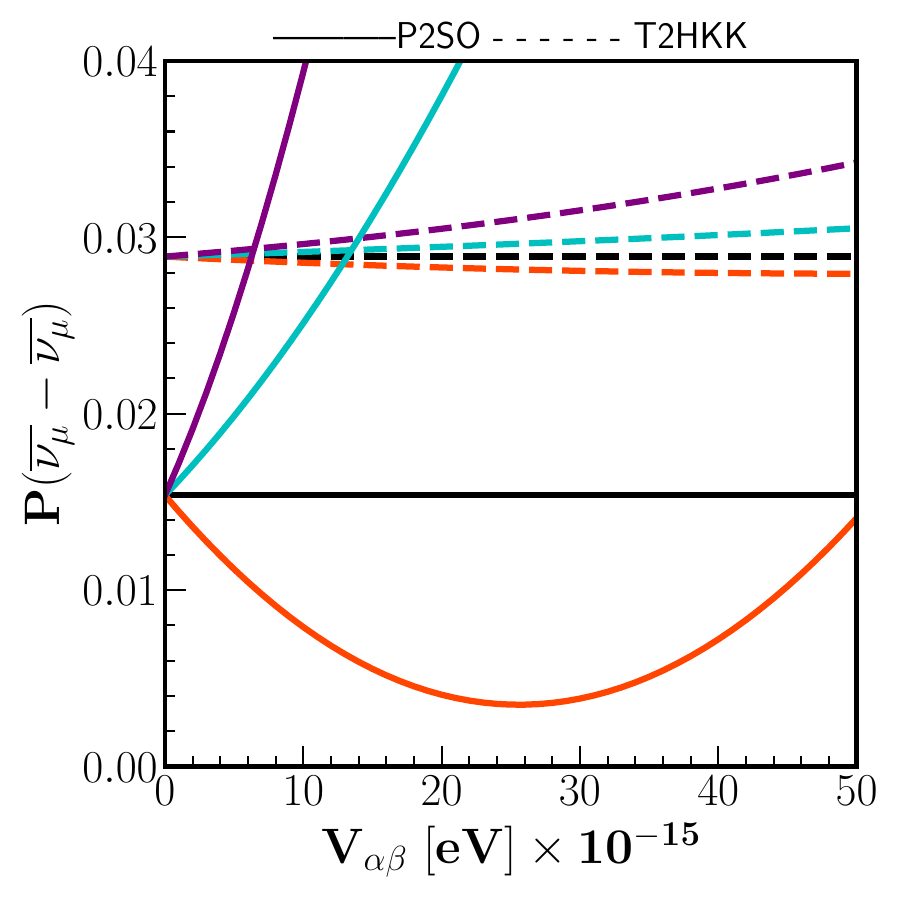}  
    \caption{Appearance and disappearance probabilities for neutrino and antineutrino cases as a function of LRF potentials for P2SO and T2HKK. The value of $\deltaCP$ is taken to be zero for all cases and values of other oscillation paeameters is taken from Tab. \ref{table_sparam}.  Neutrino energy are taken  where the neutrino events peak.}
    \label{fig:probability}
    \end{center}
\end{figure}
\begin{figure}
\begin{center}
    \includegraphics[width=80mm, height=70mm]{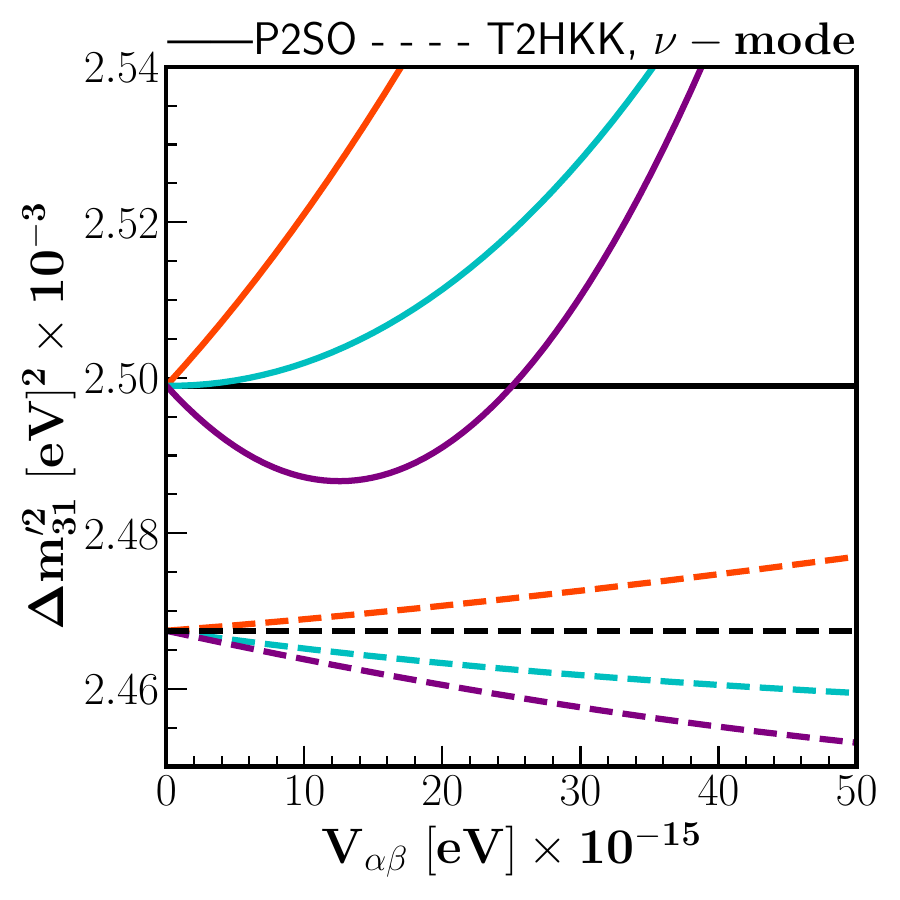}
    \includegraphics[width=80mm, height=70mm]{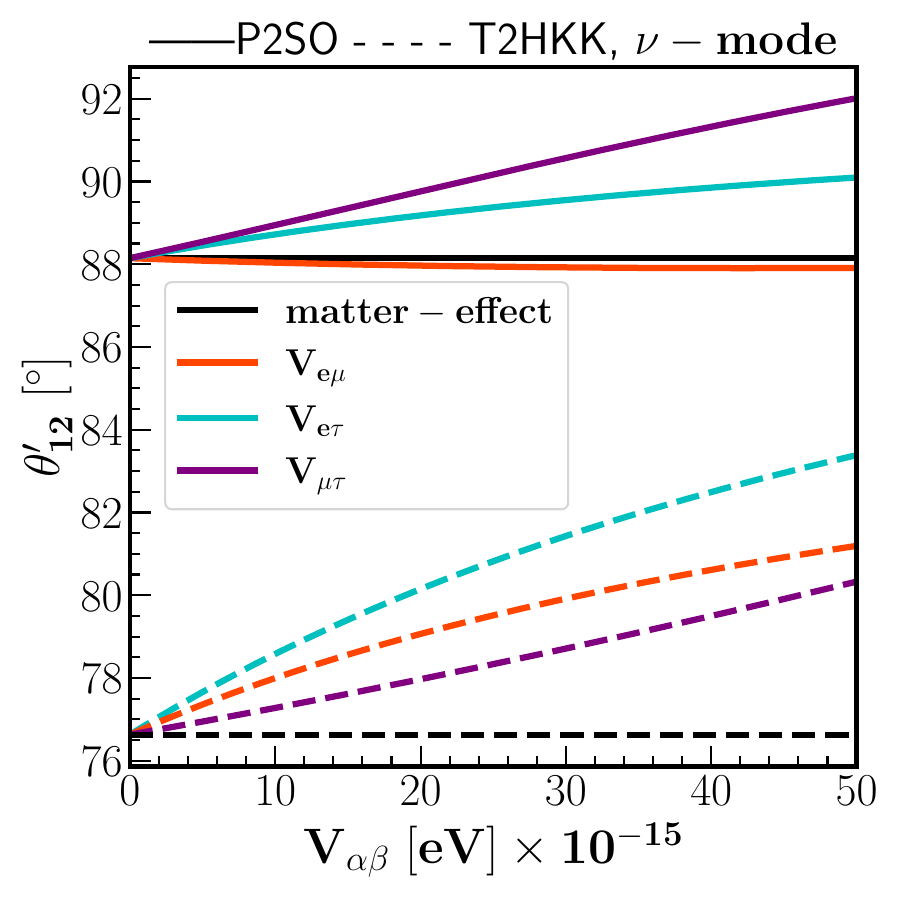}
    \includegraphics[width=80mm, height=70mm]{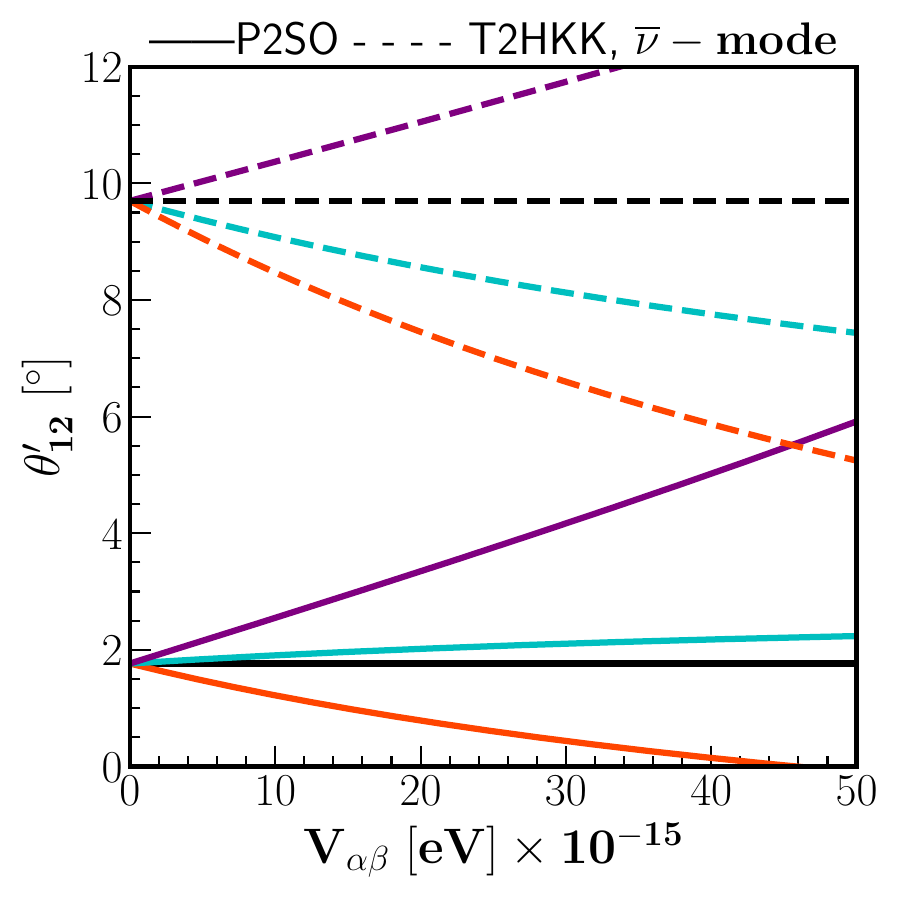}

    \caption{Dependence of  the relevant modified oscillation parameters on the LRF potential $V_{\alpha\beta}$, which are mainly responsible for the oscillation probabilities presented in Fig.~\ref{fig:probability}.}
    \label{fig:effective-para}
    \end{center}
\end{figure}


\subsection{Effect on the probabilities}
First, let us examine the impact of LRF on the oscillation probabilities in long-baseline experiments. Fig.~\ref{fig:probability} illustrates the probabilities for Standard Interaction (SI) and LRF potentials in P2SO and T2HKK. The left and right panels represent the appearance and disappearance probabilities,  while the upper and lower plots depict the neutrino and antineutrino cases, respectively. In each panel, the black line denotes the probability in the SI case. Solid lines correspond to P2SO, and dashed lines are for T2HKK. The orange, cyan, and purple curves indicate the variations in probabilities with respect to the LRF parameters $\vem$, $\vet$ and $\vmt$, respectively. These panels are generated for a neutrino energy of 4.9 GeV for P2SO and 0.75 GeV for T2HKK. These are the energies where the event spectrum peaks in neutrino modes  for these experiments. Here it is important to note that for the T2HKK baseline, the first oscillation maximum does not occur at 0.75 GeV. Rather, this energy corresponds somewhere between the first and the second oscillation maximum. For this reason, the values of the neutrino probabilities and the antineutrino probabilities for T2HKK are  not very different at $V_{\alpha \beta} = 0$. Whereas for P2SO, at $V_{\alpha \beta} = 0$, the neutrino probabilities are higher than antineutrino probabilities as for this baseline, 4.9 GeV is close to the first oscillation maximum. From the figure, we see that for both P2SO and T2HKK, LRF parameters of $e\tau$ and $\mu\tau$ sectors  affect  the appearance probability significantly in neutrino case, whereas in the antineutrino appearance probability, appreciable  change occurs due to $\vem$. In the disappearance channel, all the three LRF parameters significantly affect the probabilities for P2SO in both neutrino and antineutrino cases. However, for T2HKK, the effects of $V_{e \mu}$ is higher for neutrinos only and the effect of $V_{\mu \tau}$ is higher for both neutrinos and   antineutrinos. From the panels, we also see certain differences in the probabilities for P2SO and T2HKK. For example, in the top left panel, the probabilities are increasing functions of $V_{e\tau}$ and $V_{\mu\tau}$ for P2SO, but they are decreasing functions in T2HKK. Some differences in the nature of the probability curves in P2SO and T2HKK can also be observed in the disappearance channel probabilities for both neutrinos and antineutrinos. To understand these differences, one way is to look at the effective values of the mixing parameters in presence of LRF, which are shown in Fig.~\ref{fig:effective-para}.

\begin{figure}[h!]
\begin{center}
    \includegraphics[width=162mm, height=64mm]{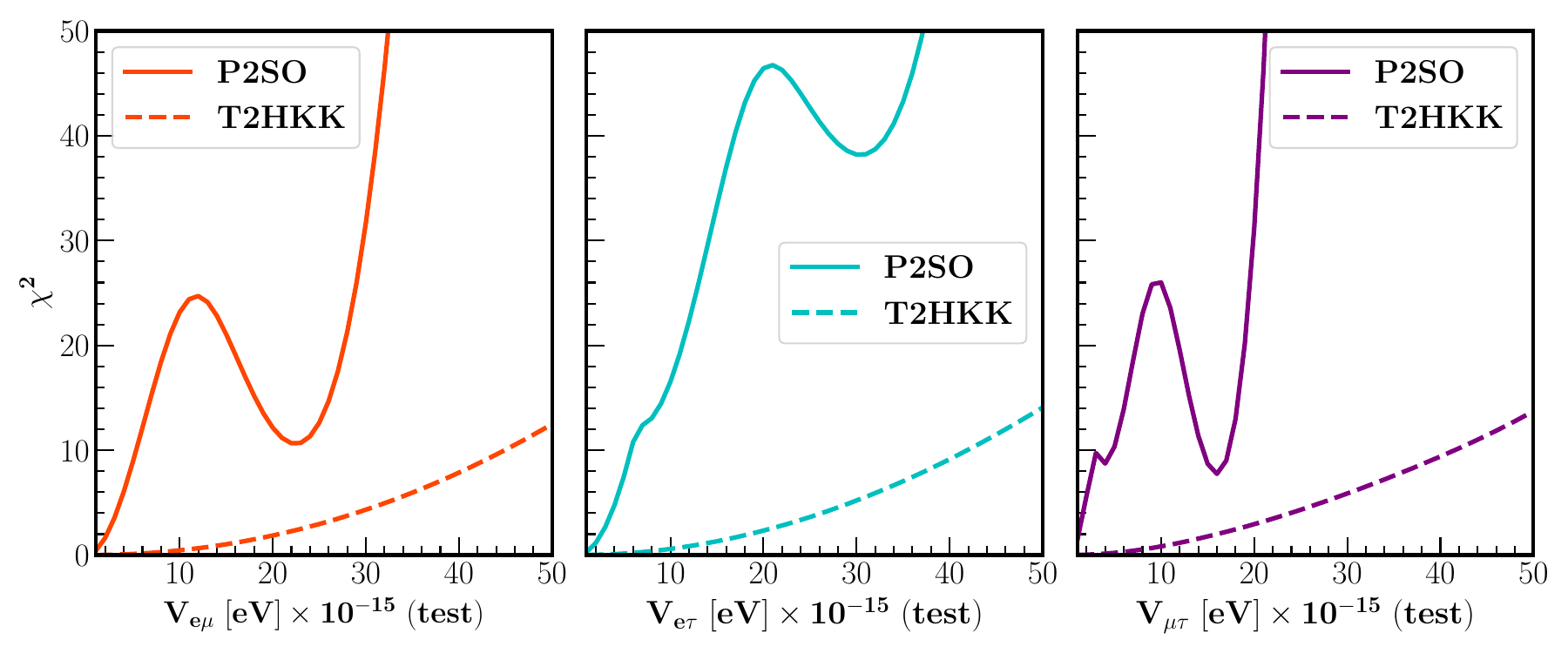}
        \caption{Sensitivity limits on LRF parameters $V_{e\mu}$,  $V_{e\tau}$, and $V_{\mu\tau}$ for P2SO and T2HKK experiments. The values of all oscillation parameters are taken from Tab. \ref{table_sparam}.}
    \label{fig:bound-P2SO-T2HKK}
    \end{center}
\end{figure}

\begin{figure}[htbp!]
\begin{center}  
\includegraphics[width=162mm, height=64mm]{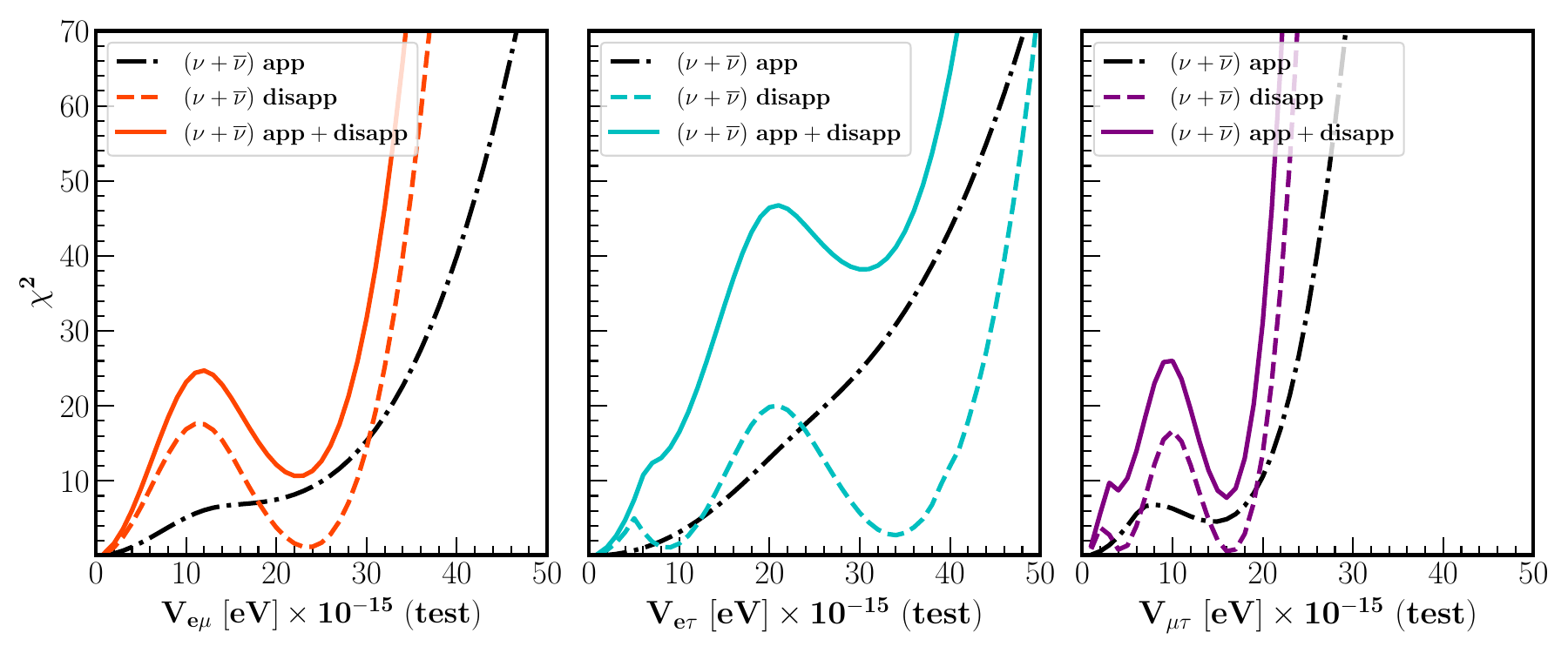}\\
    \caption{Sensitivity limits on LRF parameters $V_{e\mu}$,  $V_{e\tau}$, and $V_{\mu\tau}$ in appearance, disappearance and in combined case for P2SO experiment, for $\deltaCP(\rm true) = 232^{\circ}$.}
    \label{fig:Bound}
    \end{center}
\end{figure}
\begin{figure}[htbp!]
\begin{center}  
\includegraphics[width=162mm, height=64mm]{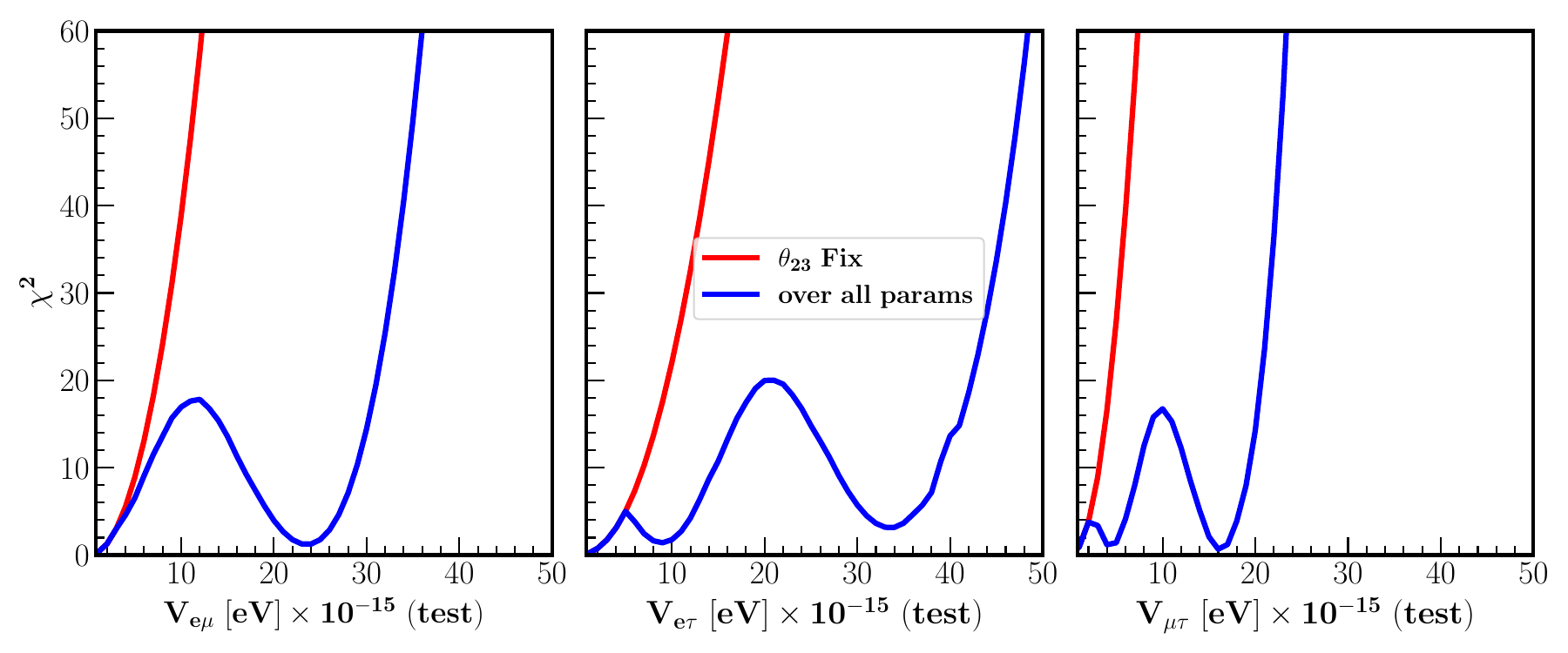}\\
    \caption{ Sensitivity limits on LRF parameters in the disappearance channel. In each of the three panels, the red curve illustrates the sensitivity limits of LRF parameters while maintaining $\theta_{23}$ value fixed at both true and test values . Conversely, the blue curve depicts the sensitivity bounds of LRF parameters when the true values of oscillation parameters, as specified in Tab.~\ref{table_sparam}, are retained, and marginalization is performed across all parameters.}
    \label{fig:Bound-new}
    \end{center}
\end{figure}

In the top left, top right and bottom panels of Fig.~\ref{fig:effective-para}, we present the variation of the effective parameters $\Delta m^{\prime 2}_{31}$ for neutrinos, $\theta^\prime_{12}$ for neutrinos and $\theta^\prime_{12}$ for antineutrinos, respectively, as a function of the LRF parameter $V_{\alpha \beta}$. The other specifications of these panels are exactly same as Fig.~\ref{fig:probability}. From these plots, we see that for some of these curves, the behaviour in P2SO and T2HKK are quite different. For example, in the top left panel, the cyan curve is an increasing function of $V_{e\tau}$ for P2SO, whereas it is a decreasing function of $V_{e\tau}$ for T2HKK. Further, the nature of the purple curve is also different for P2SO and T2HKK. Similarly, the differences between P2SO and T2HKK can also be seen in the orange curve in the top right panel and the cyan curve in the bottom panel. Because of these differences in the behaviour of the effective mixing parameters in P2SO and T2HKK, the probability curves also become different for these two experiments. We have checked that, the behaviour of the other effective mixing parameters, which are not shown for Fig.~\ref{fig:effective-para} are similar in P2SO and T2HKK.

\subsection{Sensitivity limits on the LRF parameters}

\begin{figure}[htbp!]
\begin{center}  
\includegraphics[width=162mm, height=64mm]{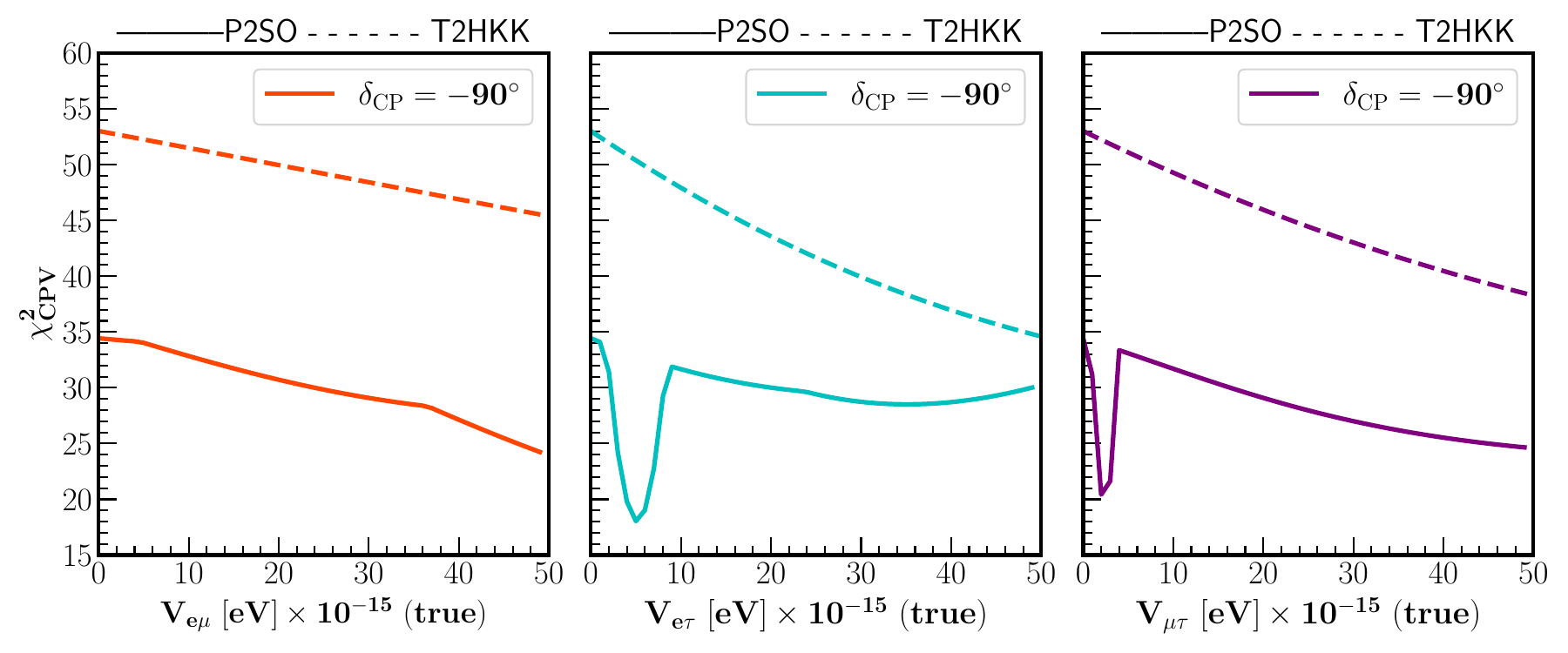}\\
\caption{CPV sensitivities as a function of $V_{\alpha\beta}$ for the true value of $\deltaCP = -90^\circ$ for P2SO and T2HKK experiments.}
    \label{fig:CPV-sensitivity}
    \end{center}
\end{figure}

\begin{figure}[htbp!]
    \centering
    \includegraphics[width=162mm, height=64mm]{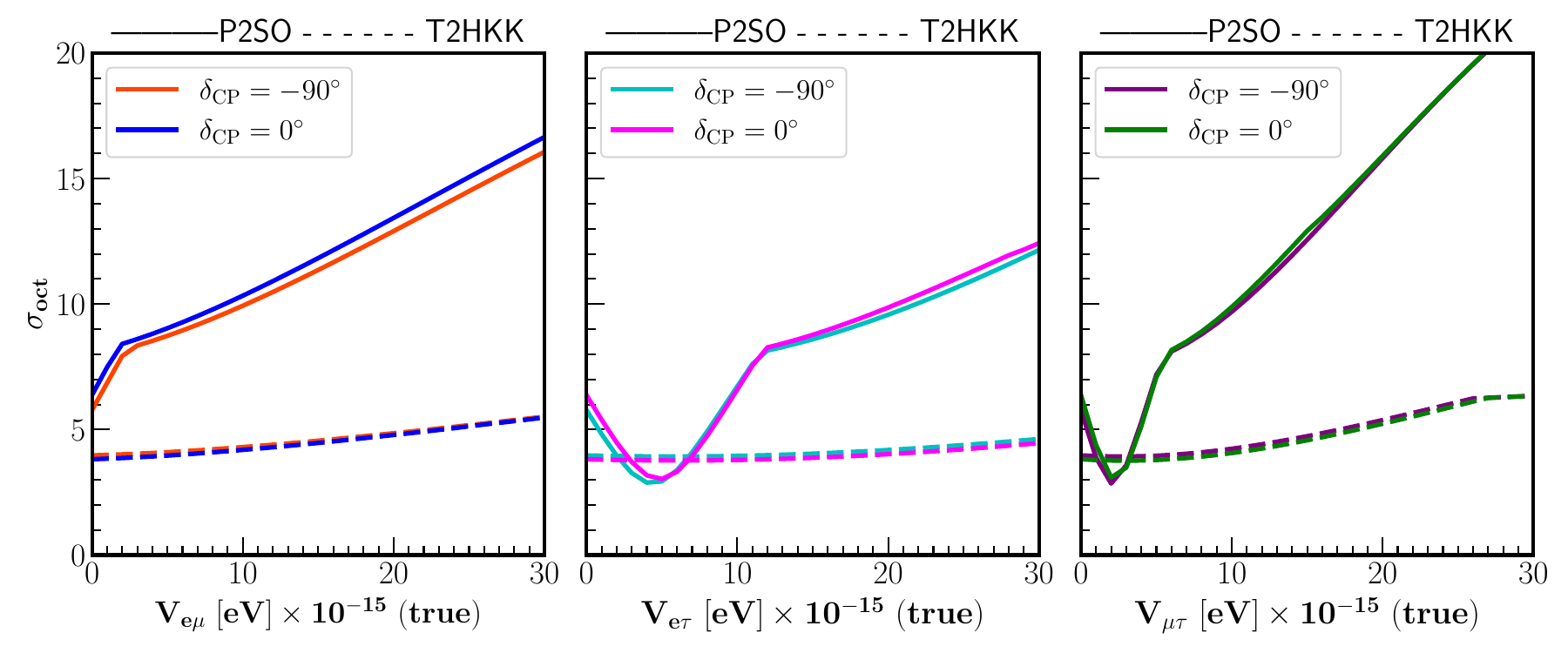}    
    \caption{Octant sensitivity as a function of $V_{\alpha\beta}$ for two different true values of $\deltaCP$ for P2SO and T2HKK experiments. We have considered the parameter $\theta_{23}$ to be LO for the analysis.}
    \label{fig:Octant-sensitivities}
\end{figure}

In this section, we study the capability of P2SO and T2HKK to put limits on the LRF parameters. In Fig.~\ref{fig:bound-P2SO-T2HKK}, we show the bounds on the LRF parameters corresponding to $e\mu$, $e\tau$, and $\mu\tau$ sectors from P2SO and T2HKK experiments. This has been obtained by taking the standard scenario in the true spectrum and the LRF scenario in the test spectrum for the calculation of $\chi^2$. Left, middle, and right panels are for $V_{e\mu}$,  $V_{e\tau}$, and $V_{\mu\tau}$, respectively. In each panel solid curve represents the sensitivity limit for P2SO and dashed curve is for T2HKK experiment. More constrained bounds are obtained from P2SO experiment as compared to T2HKK for all the LRF parameters due to the longer baseline and higher statistics of P2SO experiment. Sensitivity limits on the LRF parameters are shown in Tab.~\ref{tab:bound-table} at 90$\%$ C.L. from P2SO and T2HKK. In that table, we also list the bounds from Super-Kamiokande (SK), INO, DUNE and T2HK experiments in order to compare our results with the bounds form other experiments. From the table, we see that P2SO experiment can put the strongest bounds on the LRF parameters. The bounds obtained from T2HKK are better than T2HK but weaker than DUNE. The bounds obtained from SK are somewhat poor and the future atmospheric experiment INO is expected to put stronger bounds on the LRF parameters than T2HKK. 

In Fig.~\ref{fig:bound-P2SO-T2HKK}, we see a dip in the sensitivity curves for P2SO in all three panels. However, these dips are not present in the T2HKK curves. To understand this, in Fig.~\ref{fig:Bound}, we show the contribution from the individual probability channels for the P2SO experiment. In each panel solid curve is for the combination of appearance and disappearance events. Dashed (dash-dotted) curves in each panel are the sensitivities considering only disappearance (appearance) events. It is clear from the panels that, the dip is mainly due to the contribution of disappearance neutrino events. 

To understand the origin of the dip coming from the disappearance channel, we tried to look into the effect of marginalization. In Fig.~\ref{fig:Bound-new}, we show how the sensitivity of the disappearance channel changes with respect to the marginalization of the parameter $\theta_{23}$. From the panels, we see that when $\theta_{23}$ is kept fixed in the test, the dip disappears in all three cases. This implies the presence of octant degeneracy (i.e., $\theta_{23} \rightarrow 90^\circ - \theta_{23}$) in the disappearance channel of P2SO. As this degeneracy is not present for T2HKK, we do not see any dip for this experiment.

\begin{table}
    \centering
    \begin{tabular}{|c||c|c|c|c|c|c|}
    \hline
        \textbf{LRF Potential $[eV] $} & ~~\textbf{SK} \cite{Joshipura:2003jh} ~~&~~\textbf{INO} \cite{Khatun:2018lzs}~~&~ \textbf{DUNE} \cite{Singh:2023nek} ~& ~ \textbf{T2HK} \cite{Singh:2023nek} ~&~ \textbf{P2SO} ~&~ \textbf{T2HKK} \\
        &&&&&(This work)&(This work)\\ \hline \hline
        $\vem (\times 10^{-14})$& 71.5 & 1.56  & 1.46 & 3.45 & 0.23 & 2.40 \\
         $\vet (\times 10^{-14})$& 83.2 & 1.56 & 1.03 &  3.43& 0.23 & 2.15\\
       $\vmt (\times 10^{-14})$ & - & - & 0.67 & 1.84 & 0.13 & 1.5\\ \hline 
        \end{tabular}
    \caption{Sensitivity limits at $90\%$ C.L. on LRF parameters from several experiments.}
    \label{tab:bound-table}
\end{table}


\subsection{Effects of LRF parameters on CP violation, Octant and mass ordering sensitivities}

\begin{figure}[htbp!]
\begin{center}
    \includegraphics[width=162mm, height=64mm]{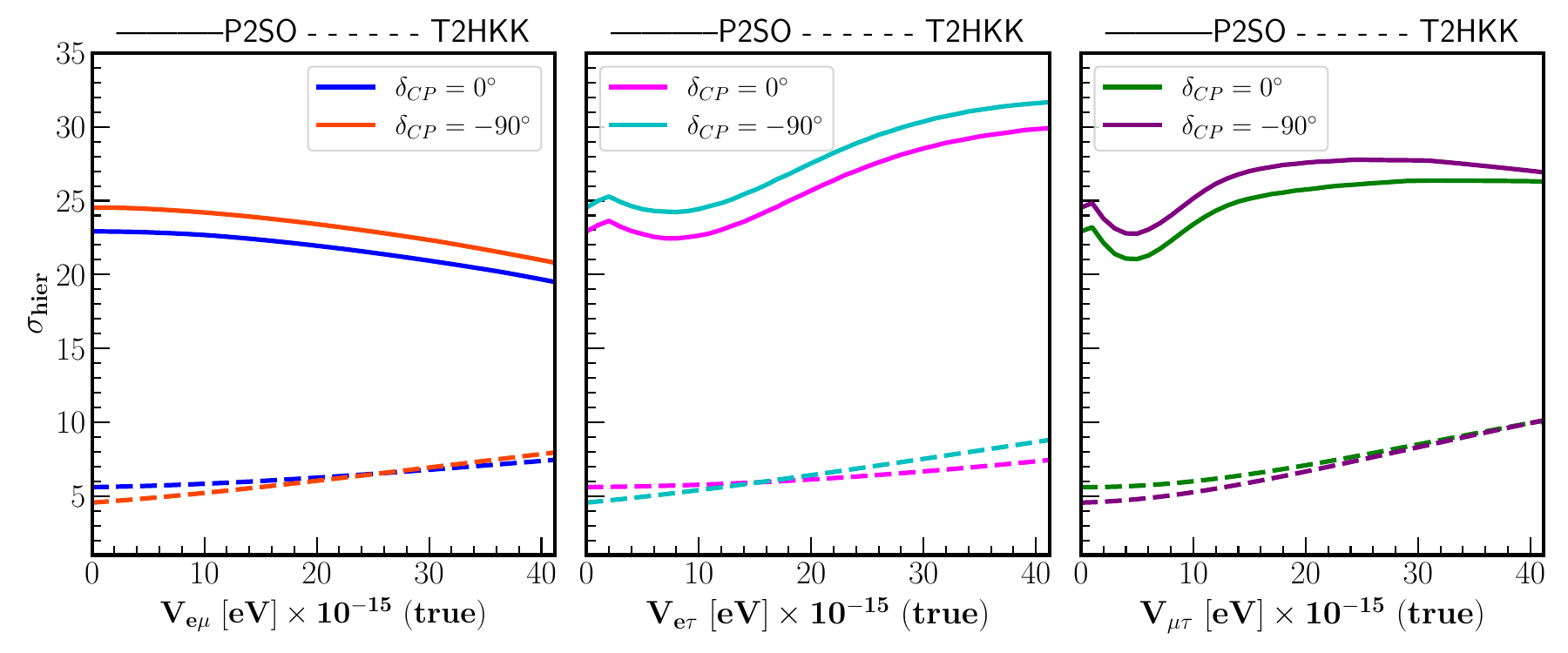}  
    \caption{Variation of neutrino mass ordering sensitivities with respect to $V_{\alpha\beta}$ for both T2HKK and P2SO experiments.}
    \label{fig:MH-sensitivities}
    \end{center}
\end{figure}

In this section, we discuss how LRF can affect the sensitivities of P2SO and T2HKK in determining the unknowns in the neutrino oscillation. We do this by keeping $V_{\alpha \beta}$ fixed in both true and test spectra of  $\chi^2$. At present, there are three unknowns in  neutrino oscillation formalism, which are: (i) true ordering of the neutrino masses, which can be either normal or inverted, (ii) true octant of $\theta_{23}$ which can be lower or higher and (iii) the value of $\delta_{\rm CP}$ which can lead to CP violation (CPV) in the neutrino sector. 
Let us begin with the CPV sensitivity. Fig.~\ref{fig:CPV-sensitivity} shows the CPV sensitivity as a function of $V_{\alpha\beta}$. Left/middle/right panel shows the sensitivity for $\vem / \vet / \vmt$. We have obtained CPV sensitivity by excluding CP conserving values of $\deltaCP$ for true value of $\deltaCP$ as $-90^\circ$. Solid curves represent the sensitivities for P2SO while dashed curves are for T2HKK experiment. CPV sensitivities of T2HKK are higher compared to P2SO experiment for all the LRF parameters as well as SI. In general, we see that as $V_{\alpha \beta}$ increases, the CPV sensitivity decreases. Additionally, we observe kinks in the $\vet$ and $\vmt$ curves. We have checked that these kinks appear due to the degeneracy associated with the parameter $\theta_{23}$. If we keep $\theta_{23}$ fixed in our calculation, these kinks disappear.

\begin{figure}
    \centering
\includegraphics[width=162mm, height=64mm]{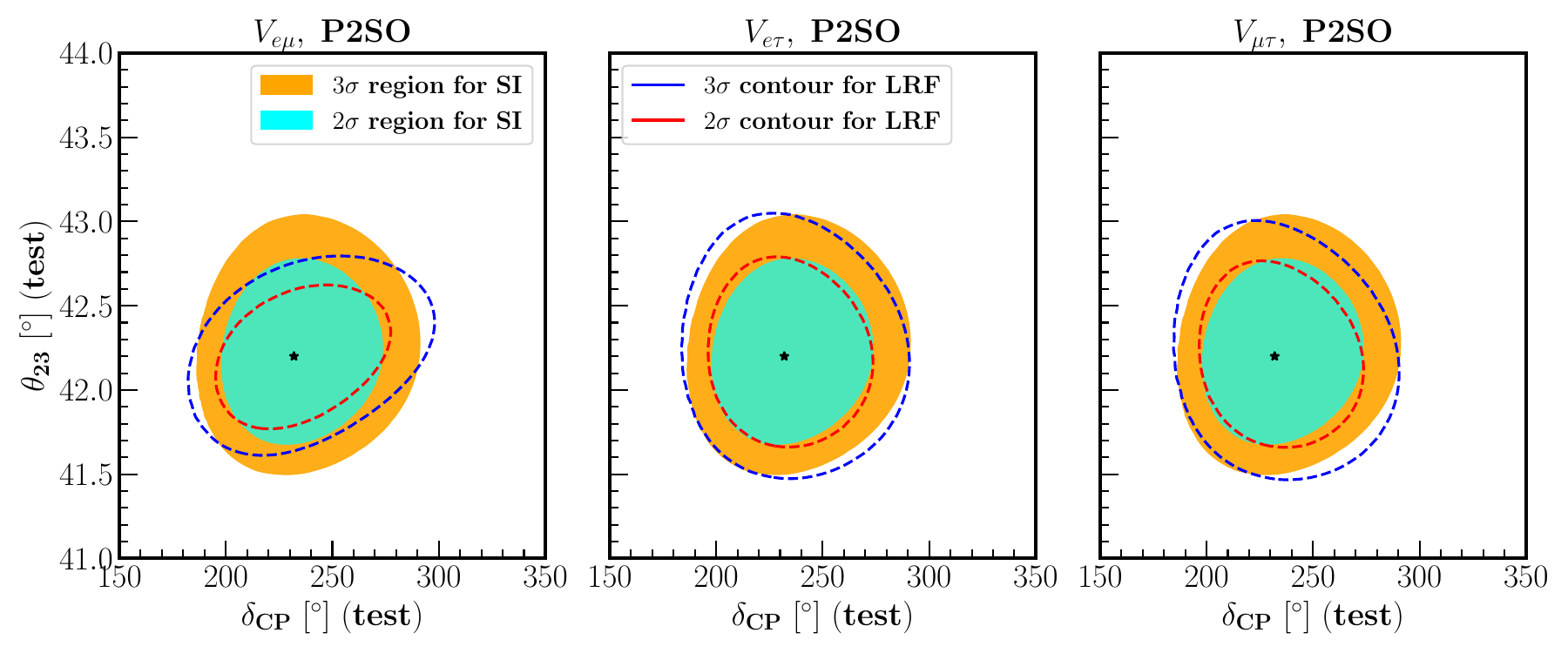}
\includegraphics[width=162mm, height=64mm]{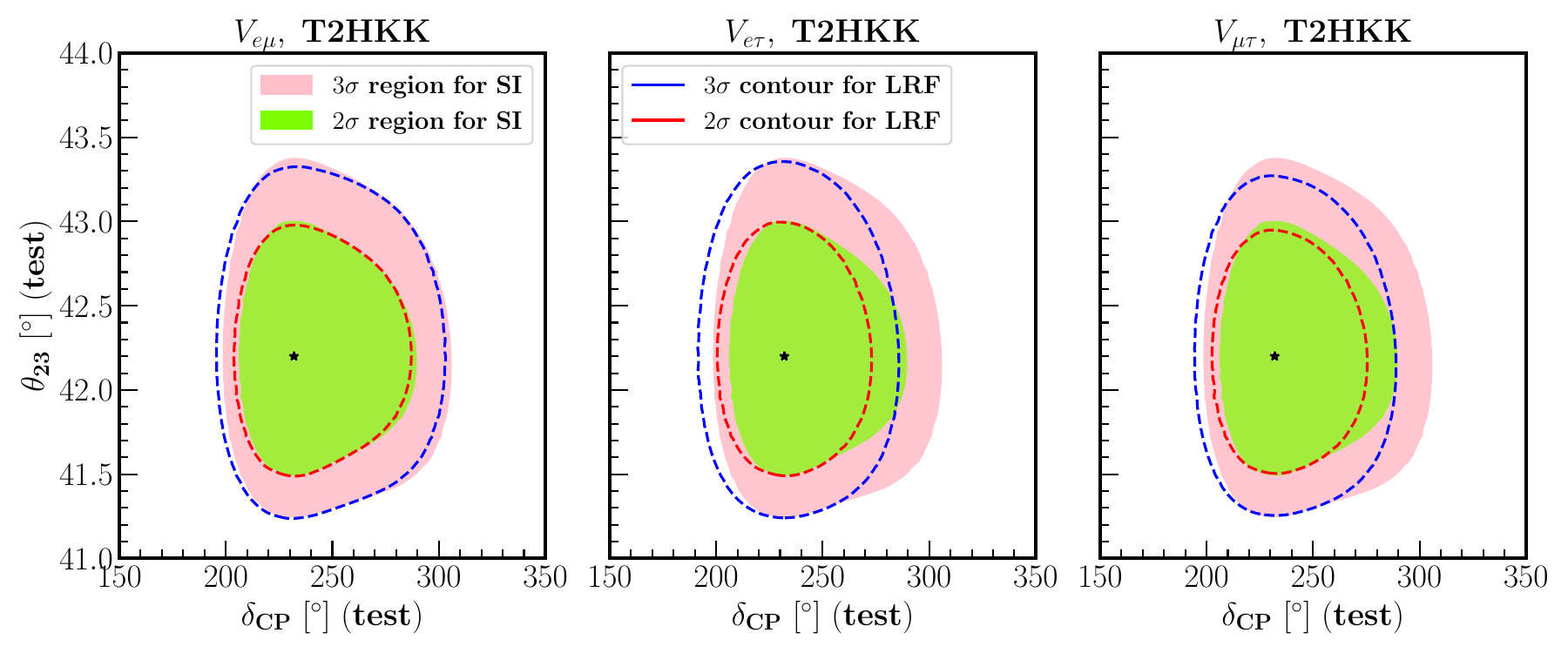}
    \caption{Allowed parameter space between $\theta_{23}$ (test) $\delta_{\rm CP}$ (test) with $V_{\alpha \beta}$ fixed to their upper bound values at 5 $\sigma$ C.L. for P2SO and T2HKK.}
    \label{fig:dcp-th23-5sig-p2so}
\end{figure}

In Fig.~\ref{fig:Octant-sensitivities}, we show the same as Fig.~\ref{fig:CPV-sensitivity} but for octant sensitivity. We obtain the octant sensitivity by considering the true $ \theta_{23}$ in lower octant (LO)  and varying the test $\theta_{23}$ in higher octant (HO). We show this for two values of $\delta_{\rm CP}$ i.e., $0^\circ$ and $-90^\circ$. From the panels we see that, in general, octant sensitivity is higher in P2SO as compared to T2HKK.  Further, we see that sensitivity is almost similar for both the values of  $\delta_{\rm CP}$. For T2HKK, we see that octant sensitivity continuously increases as $V_{\alpha \beta}$ increases. Whereas for P2SO, octant sensitivity increases continuously for $V_{e\mu}$ and for the other two $V_{e\tau}$ and $V_{\mu\tau}$, the sensitivity curve has a dip. We have checked that this dip occurs because the $\chi^2$ minimum appears with different values of $\theta_{23}$ for different $V_{\alpha \beta}$. The change of sensitivity with respect to $V_{\alpha \beta}$ is much higher in P2SO as compared to T2HKK.  

Next, we study the effect of LRF parameters on the neutrino mass ordering sensitivity. Mass ordering sensitivity signifies the ability of the experiment to determine the true ordering of the neutrino masses. Fig.~\ref{fig:MH-sensitivities} shows the mass ordering sensitivities as a function of $V_{\alpha\beta}$ for two different true values of $\deltaCP$ i.e., $0^\circ $ and  $-90^\circ$. The labelling of the panels is  same as Figs.~\ref{fig:CPV-sensitivity} and \ref{fig:Octant-sensitivities}. Here also, we see that the sensitivity of P2SO is higher than T2HKK and sensitivities are similar for both the values of $\deltaCP$. However, it is interesting to see that for $V_{e\mu}$, the sensitivity increases for T2HKK and decreases for P2SO as $V_{e\mu}$ increases. For the other two $V_{\alpha \beta}$, sensitivity increases for both the experiments as $V_{\alpha \beta}$ increases. We also observe a kink in P2SO for $\vmt$ and $\vet$. We have verified that these kinks are due to the large backgrounds of the P2SO experiment \cite{Singha:2021jkn}.

 Here we observe that P2SO has better sensitivity for mass ordering and octant as compared to T2HKK whereas for CPV, T2HKK performs better. In general, the mass ordering and octant sensitivities depend on the matter effect whereas the CP sensitivity depends on the statistics. Further, the experiment P2SO is sensitive towards the first oscillation maximum whereas T2HKK probes both the first and second oscillation maxima. While the first maximum corresponds to a higher value of $E$ giving rise to enhanced matter effect, the CPV sensitivity is higher at the second oscillation maximum \cite{Parke:2013pna}. The P2SO experiment has larger matter effect due to longer-baseline as well as the energy of the neutrinos at the first oscillation maximum. For this reason it is more sensitive to mass ordering and octant in spite of the fact that its sensitivity is limited by background \cite{Singha:2021jkn}. Whereas in T2HKK the CPV is enhanced due to the effect of the second oscillation maximum as well as the shorter baseline giving large statistics.


\subsection{Effect of LRF parameters in the precision of $\deltaCP$, $\Delta m^2_{31}$ and $\theta_{23}$}
\label{correlation}

\begin{figure}
    \centering
    \includegraphics[width=162mm, height=64mm]{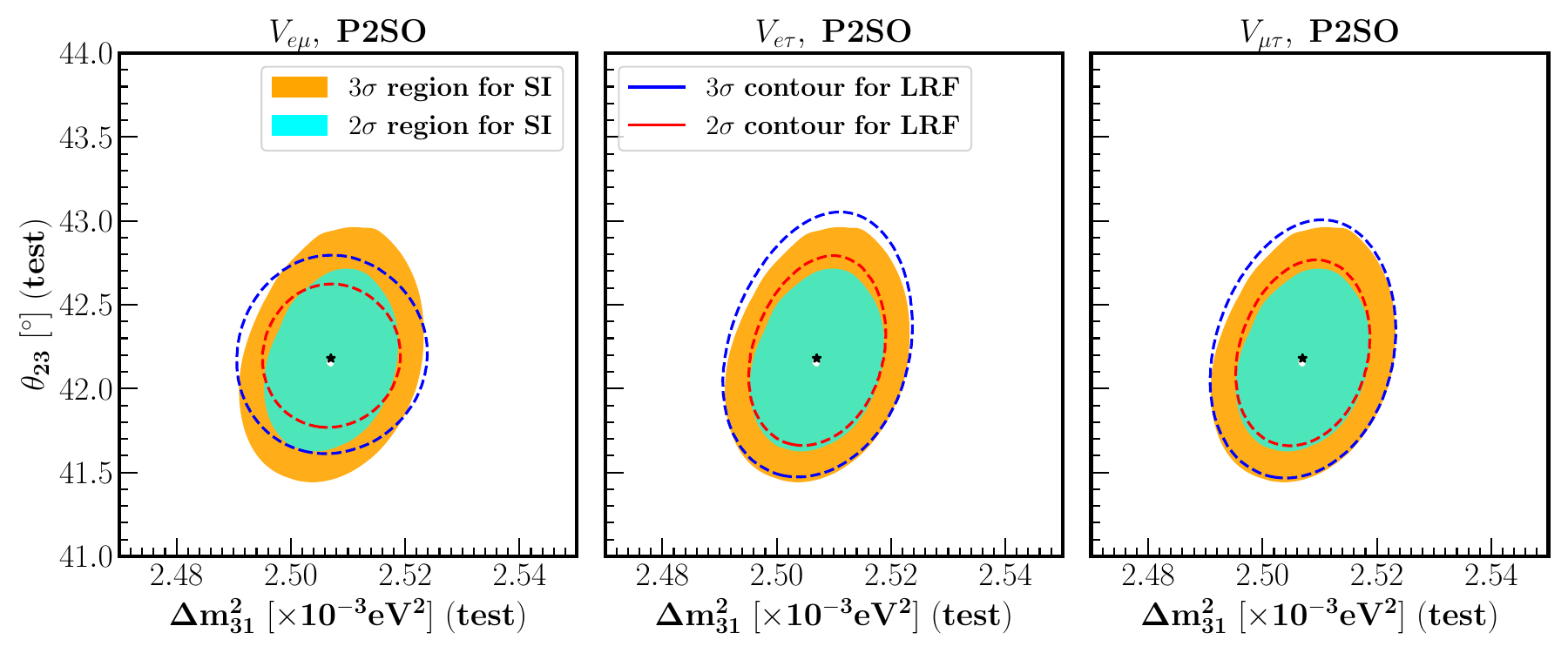}
    \includegraphics[width=162mm, height=64mm]{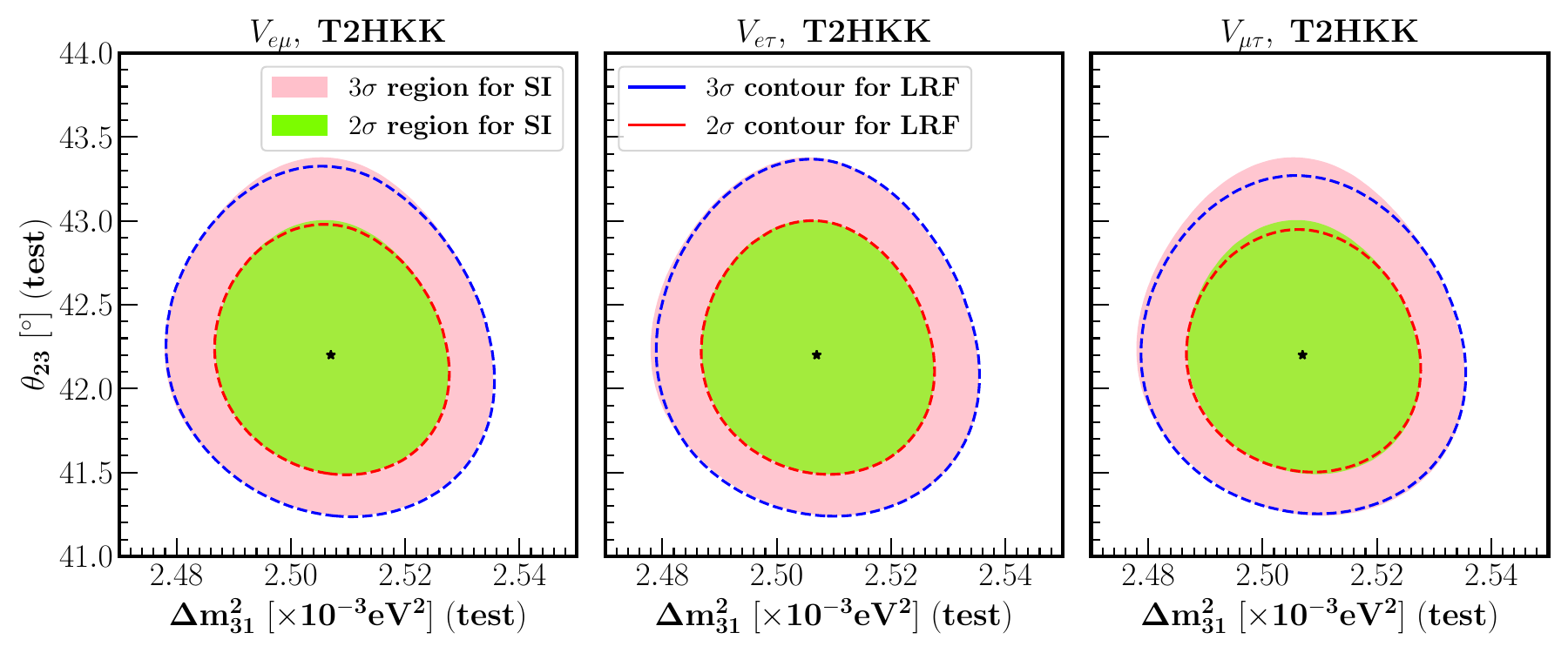}
    \caption{Allowed parameter space between $\theta_{23}$ (test) $\Delta m^2_{31}$ (test) with $V_{\alpha \beta}$ fixed to their upper bound values at 5 $\sigma$ C.L. for P2SO and T2HKK.}
    \label{fig:ldm-th23-3sig-p2so}
\end{figure}

In this section, we study the effect of the LRF parameters in the precision measurements of $\deltaCP$, $\Delta m^2_{31}$ and $\theta_{23}$.
To show this, we have kept $V_{\alpha \beta}$ fixed in both true and test spectra of the $\chi^2$ using their $5 \sigma$ upper limit values. Then, taking the current true values of $\deltaCP$, $\Delta m^2_{31}$ and $\theta_{23}$, we show 2-D allowed regions of these parameters at $2 \sigma$ C.L. and $3 \sigma$ C.L. 

Fig.~\ref{fig:dcp-th23-5sig-p2so} shows the allowed regions in $\deltaCP - \theta_{23}$ plane. The top row is for P2SO and the bottom row is for T2HKK. In each row, left, middle and right panels are for $V_{e \mu}$, $V_{e \tau}$ and $V_{\mu\tau}$ respectively. In each panel, dashed red (blue) curve is 2$\sigma$ (3$\sigma$) contours in the presence of LRF and the shaded regions  represent the SI for the respective experiments. Asterisk symbol in each panel represents the true values of corresponding oscillation parameters. From these panels, we note certain changes in the precision of $\deltaCP$ and $\theta_{23}$ in presence of LRF. For T2HKK, the precision of $\theta_{23}$, almost remains same as in the SI case for all $V_{\alpha \beta}$. Whereas for  $\delta_{\rm CP}$, the precision improves as compared to SI case for $V_{e\tau}$ and $V_{\mu\tau}$. For P2SO, the precision to both $\deltaCP$ and $\theta_{23}$ are almost similar to the SI case for $V_{e\tau}$ and $V_{\mu\tau}$. For $V_{e\mu}$, the precision of $\theta_{23}$ is better than the SI  and the precision of $\deltaCP$ is slightly poor than the SI scenario. 

To show the effect of LRF parameter on the precision of $\Delta m^2_{31}$, we adopt the same approach for generating the allowed parameter space in $\Delta m^2_{31}-\theta_{23}$ plane. The results are presented in Fig.~\ref{fig:ldm-th23-3sig-p2so}. From the panels, we see that the precision of $\Delta m^2_{31}$ remains almost unaltered as compared to SI scenario in the presence of LRF.


\subsection{Sensitivity limits on $M_{Z_{\alpha \beta}}$ and $g_{\alpha \beta}$}

\begin{figure}
    \centering
    \includegraphics[width=162mm, height=64mm]{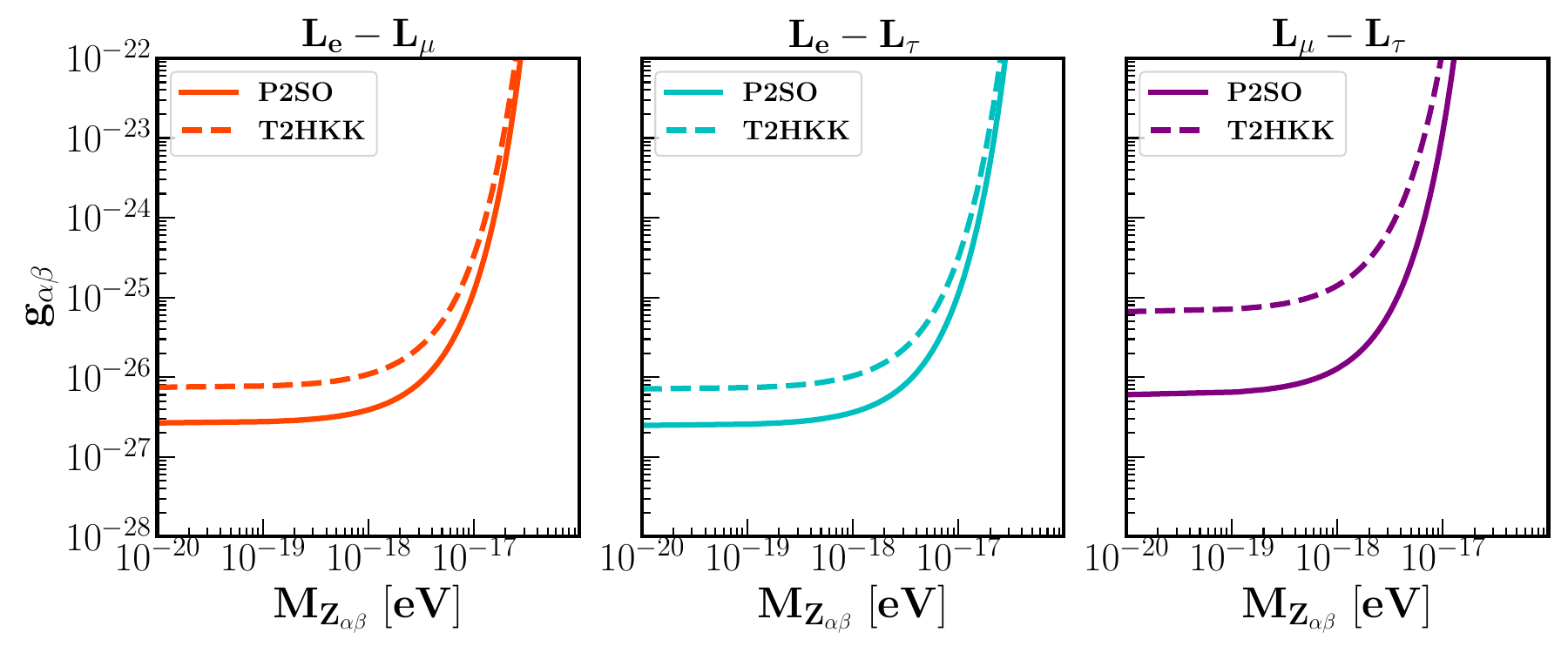}
    \caption{Plots portraying the allowed range for gauge coupling vs mass of gauge boson for the long range potential in all three cases, is at $2\sigma$ C.L. for both P2SO and T2HKK. For more details please refer to the Sec. \ref{correlation}}
    \label{fig:M-gz}
\end{figure}

\begin{table} 
 \begin{center}
\begin{tabular}{|l||*{5}{c|}}\hline
\backslashbox{Experiment}{Model}
&\makebox[10em]{$L_e-L_\mu$}&\makebox[10em]{$L_e-L_\tau$}&\makebox[10em]{$L_\mu-L_\tau$}
\\\hline\hline
 \hspace{0.25cm}P2SO (This work) & $2.66 \times 10^{-27}$ & $2.48 \times 10^{-27}$ & $6.03 \times 10^{-27}$\\\hline
 \hspace{0.25cm}T2HKK (This work) & $7.47 \times 10^{-27}$ & $7.12 \times 10^{-27}$ & $6.637 \times 10^{-26}$\\\hline
\hspace{0.25cm}T2HK \cite{Singh:2023nek} & $1.30 \times 10^{-26}$ & $1.24 \times 10^{-26}$ & $4.31 \times 10^{-26}$\\\hline
\hspace{0.25cm} DUNE \cite{Singh:2023nek} & $8.55 \times 10^{-27}$ & $7.03 \times 10^{-27}$ & $2.59 \times 10^{-26}$\\\hline
\end{tabular}
 \caption{Projected upper bound on $g_{\alpha \beta}$ from various long-baseline experiments.}
 \label{Tab:Z-bound}
 \end{center}
 \end{table}

Neutrino oscillation experiments can also give us the opportunity to put bounds on the mass of the new gauge boson $M_{Z_{\alpha \beta}}$ and the new gauge coupling $g_{\alpha \beta}$. In Fig.~\ref{fig:M-gz}, we present the same at $2\sigma$ C.L., corresponding to the long-range force experienced by the neutrinos at Earth due to the matter density of the Sun, using Eqs.~\eqref{v_ej} and \eqref{v_mt}. In each panel, the dashed curve is for T2HKK while the solid curve portrays the allowed region for P2SO. From the panels, we see that the bounds obtained for P2SO are better than T2HKK and among the three symmetries, the most stringent bound comes from $L_e - L_\tau$ gauge symmetry. From these plots, we find the upper bound for the coupling of ultra light mediators with mass $M_{Z_{\alpha \beta}}\lesssim 10^{-17}$ eV to be $g_{\alpha \beta} \lesssim 10^{-26}$ for $L_\alpha-L_{\beta}$ models.
Earlier bounds on $g_{\alpha \beta}$ using Super-Kamiokande data are: $g_{e \mu}< 8.32 \times 10^{-26}$ and $g_{e \tau}< 8.97 \times 10^{-26}$ at $90\%$ confidence level \cite{Joshipura:2003jh}. Using solar and reactor data from KamLAND, the bounds obtained in Ref. \cite{Bandyopadhyay:2006uh} are $g_{e \mu}< 2.06 \times 10^{-26}$ and $g_{e \tau}< 1.77 \times 10^{-26}$, at $3 \sigma$, assuming $\theta_{13}=0^\circ$. The projected bounds on $g_{\alpha \beta}$ for the upcoming DUNE and T2HK experiments are obtained in \cite{Singh:2023nek}. The limit on gauge coupling is also established based on the energy dissipation due to $Z_{\mu\tau}$ radiation emitted by compact binary systems, yielding $g_{\mu\tau} < 10^{-20}$ for $M_{Z_{\mu\tau}} < 10^{-19} $ eV \cite{KumarPoddar:2019ceq}. Furthermore, the constraints on $g_{e\mu}$ and $g_{e\tau}$ have been obtained in Ref. \cite{KumarPoddar:2020kdz} using the perihelion precision of planets for $M_{Z_{ej}} < 10^{-19} $ eV,  as $g_{ej} \leq 10^{-25}$.  For comparison, we list the corresponding excluded ranges for the $Z_{\alpha \beta}$ coupling and mass obtained from different long-baseline experiments in Tab.~\ref{Tab:Z-bound}.  From the table we see that the bounds obtained from P2SO are better as compared to other future long-baseline neutrino experiments.


\section{Summary and conclusion}
\label{sum}

In this paper, we have studied the sensitivity of the two upcoming long-baseline experiments P2SO and T2HKK to the long range force. The proposed P2SO experiment will study the oscillations of the neutrinos originating from protvino and to be detected at the Super-ORCA detector. Whereas the T2HKK experiment, which is an alternative option of the T2HK experiment, will have a dual baseline configuration i.e., one detector will be located in Japan and another detector will be located in Korea. 

When the Standard Model is extended with $L_e-L_\mu$, $L_e-L_\tau$ and $L_\mu-L_\tau$ type $U(1)$ gauge symmetries, it gives rise to a new gauge boson  $Z_{\alpha \beta}$ and a new coupling constant  $g_{\alpha \beta}$. If the new mediator is inordinately lightweight, it gives rise to a long range potential which can affect the propagation of the neutrinos in the matter. The nature of this new long range  potential $V_{\alpha \beta}$ would be different depending on the symmetry i.e., $L_e-L_\mu$, $L_e-L_\tau$ and $L_\mu-L_\tau$.  

In the first part of our paper, we have put forward a prescription on how to calculate the neutrino oscillation probabilities in the presence of LRF and studied how this new potential affects the neutrino oscillation probabilities in P2SO and T2HKK. In our study, we find that the effect of the LRF parameters $V_{e \tau}$ and $V_{\mu \tau}$ is more in the appearance probability in neutrino case, whereas in the antineutrino appearance probabilities, only significant change occurs due to $V_{e \mu}$. In the disappearance channel, all the three LRF parameters significantly affect the probabilities in P2SO in both neutrinos and antineutrinos whereas for T2HKK, the effect of $V_{e \mu}$ and $V_{\mu \tau}$ is higher in neutrinos and the effect of $V_{\mu \tau}$ is higher in antineutrinos. We have also noticed some significant difference between the T2HKK and the P2SO probabilities. This is because the effective mixing parameters behave differently at these relevant energies and baselines.

Next, we moved on to study the capability of the above mentioned experiments to constrain the LRF parameters. In our analysis, we find that sensitivity of P2SO is better than T2HKK due to the former's longer baseline and higher statistics. However, the sensitivity of P2SO suffers from octant degeneracy in the disappearance channel for higher values of $V_{\alpha \beta}$.  While comparing our results with the other experiments, we found that P2SO can give the strongest bound on the LRF parameters among the current and future neutrino oscillation experiments. The bounds obtained from T2HKK are better than T2HK but poor than DUNE. The bounds obtained from SK are somewhat poor, and the future atmospheric experiment INO is expected to put stronger bounds on the LRF parameters than T2HKK.

Next, we studied the effect of LRF in the measurement of standard oscillation parameters. Regarding CP violation sensitivity, we find that the sensitivity is a decreasing function of $V_{\alpha \beta}$, whereas for octant sensitivity it is an increasing function of $V_{\alpha \beta}$ for both P2SO and T2HKK. Regarding octant sensitivity, the change of sensitivity with respect to $V_{\alpha \beta}$ is much higher in P2SO as compared to T2HKK. For mass ordering, the sensitivity is an increasing function of $V_{e \mu}$ for T2HKK but it is a decreasing function for P2SO. For the other two $V_{\alpha \beta}$, mass ordering sensitivity is an increasing function with respect to $V_{\alpha \beta}$ for both the experiments. Additionally, we noticed that the sensitivity of P2SO is affected by the degeneracy related to $\theta_{23}$ for CP violation measurement and octant measurement and it is affected by large background for the determination of mass ordering. Regarding the precision measurement of $\delta_{\rm CP}$, $\Delta m^2_{31}$ and $\theta_{23}$ we find that, LRF does not affect the precision of $\theta_{23}$ in T2HKK whereas for P2SO, the precision to both $\deltaCP$ and $\theta_{23}$ are unaltered except for $V_{e\mu}$. For $V_{e\mu}$, the precision of $\theta_{23}$ is better than the SI and the precision of $\deltaCP$ is slightly poor than the SI scenario. Additionally, for T2HKK the precision of  $\delta_{\rm CP}$ improves as compared to SI case for $V_{e\tau}$ and $V_{\mu\tau}$. The precision of $\Delta m^2_{31}$ remains unchanged in presence of LRF in both the experiments.

Finally, we have computed the bounds on the mass of the new gauge boson and its coupling strength associated with the LRF experienced by the neutrinos at Earth due to the matter density of the Sun. Our results show that the bounds obtained for P2SO are better than T2HKK and among the three symmetries, the most stringent bound comes for $L_e - L_\tau$. While comparing the bounds with the other future long-baseline experiments, we find that the best bound on these parameters comes from P2SO.

 Here we would like to emphasize that in our analysis, we used the values of the oscillation parameters from the global fit which are estimated taking the standard three flavour scenario. If one fits the data from the neutrino oscillation experiments  assuming that LRF exists in Nature, the fitted values of the oscillation parameters will change as with the inclusion of LRF,  the parameter space will also include the parameter $V$. The amount of change will depend on the value of $V$ and this will also affect our results. However, from table \ref{tab:bound-table}, we see that the current upper bound on $V$ is around $10^{-12}$ eV from the SK data and the future experiments are expected to improve this bound one or two orders of magnitude. Given the smallness of $V$, we expect the change in the oscillation parameters due to LRF will be small.

\acknowledgments

PM would like to thank Prime Minister's Research Fellows (PMRF) scheme for its financial support.  This work has been in part funded by the Ministry of Science and Education of the Republic of Croatia grant No. KK.01.1.1.01.0001.
SKP and RM  acknowledge University of Hyderabad IoE project grant no. RC1-20-012. We gratefully acknowledge the use of the CMSD HPC facility of Univ. of Hyderabad to carry out the computational work.


\bibliography{LRF-ref.bib}
\end{document}